\RequirePackage{snapshot}
\documentclass[useAMS,usenatbib]{mn2e}
\usepackage{graphicx,times, aas_macros, natbib}
\usepackage{mathrsfs, amssymb, amsmath}
\usepackage[flushleft]{threeparttable}
\usepackage{dcolumn,longtable,booktabs}
\usepackage[linewidth=1.5pt]{mdframed}
\newcolumntype{d}[1]{D{.}{.}{#1}}

\begin{document}
\title[Ram-Pressure Stripping in Massive Clusters]{Jellyfish: The origin and distribution of extreme ram-pressure stripping events in massive galaxy clusters}
\author[C.\ McPartland et al.]{Conor McPartland$^1$, Harald Ebeling$^1$, Elke Roediger$^2$ \& Kelly Blumenthal$^1$\\$^1$ Institute for Astronomy, University of Hawai'i at Manoa, 2680 Woodlawn Drive, Honolulu, HI, 96822, USA\\$^2$ E.A. Milne Centre for Astrophysics, Department of Physics \& Mathematics, University of Hull, Cottinton Road, Hull, HU6 7RX, United Kingdom}

\date{Draft version}
\maketitle

\begin{abstract}
We investigate the observational signatures and physical origin of ram-pressure stripping (RPS) in 63 massive galaxy clusters at $z=0.3-0.7$, based on images obtained with the Hubble Space Telescope.  Using a training set of a dozen ``jellyfish" galaxies identified earlier in the same imaging data, we define morphological criteria to select 211 additional, less obvious cases of  RPS.  Spectroscopic follow-up observations of 124 candidates so far confirmed 53 as cluster members.  For the brightest and most favourably aligned systems we  visually  derive estimates of the projected direction of motion based on the orientation of apparent compression shocks and debris trails.   

Our findings suggest that the onset of these events occurs primarily at large distances from the cluster core ($>400$ kpc), and that the trajectories of the affected galaxies feature high impact parameters.  Simple models show that such trajectories are highly improbable for galaxy infall along filaments but common for infall at high velocities, even after observational biases are accounted for, provided the duration of the resulting RPS events is $\lesssim$500 Myr. We thus tentatively conclude that extreme RPS events are preferentially triggered by cluster mergers, an interpretation that is supported by the disturbed dynamical state of many of the host clusters. This hypothesis implies that extreme RPS might occur also near the cores of merging poor clusters or even merging groups of galaxies.

Finally, we present nine additional ``jellyfish" galaxies at z$>$0.3 discovered by us, thereby doubling the number of such systems known at intermediate redshift.  
\end{abstract}


\section{INTRODUCTION}
Evidence of accelerated galaxy evolution in galaxy clusters has been presented as early as 1980, the most well known examples being the increased occurrence of ellipticals in dense environments \citep[i.e., the morphology-density relation;][]{dressler_1980} and the higher fraction of blue galaxies in clusters at higher redshift \citep[i.e., the Butcher-Oemler effect,][]{butcher_1984}. The physical mechanisms responsible for these effects are, however, still very much debated. A variety of processes have been proposed in the literature, ranging form slow-acting gravitational interactions such as galaxy-galaxy harassment \citep{moore_1996} to potentially extremely rapid galaxy transformations brought about by interactions with the gaseous intracluster medium (ICM).

The latter process, ram-pressure stripping (RPS) is expected to be especially efficient in massive galaxy clusters, as the pressure imparted on a galaxy is directly proportional to the local gas density of the ICM and to the square of the galaxy's velocity with respect to the ICM \citep{gunn_infall_1972}.  The resulting removal of the galaxy's interstellar medium (ISM) occurs in the direction of motion of the galaxy relative to the ICM, generating a trail of star-forming regions in the galaxy's wake.  For fortuitous viewing angles, this trail, or at least the associated deformation of the galactic disk, is accessible to observation, thus creating a rare opportunity to constrain the motion of galaxies in the plane of the sky.  Observations of RPS events thus constitute a valuable complement to spectroscopic radial-velocity surveys and permit a detailed investigation of the kinematics and spatial evolution of galaxies in the dense cluster environment.

The physics and observational signature of RPS have been the subject of extensive numerical simulations which predict that gradual stripping should be pervasive even in low-mass clusters \citep{vollmer_2001}.  Indeed RPS events have been studied in great detail in the Virgo \citep{chung_2007, vollmer_influence_2012, abramson_caught_2011} and Coma clusters \citep{smith_2010, yagi_dozen_2010}, as well as in other nearby systems, such as the Shapley Concentration \citep{merluzzi_2013} or Abell 3627 \citep{sun_2007,fumagalli_2014}.  As expected, these events are relatively modest though, with observations showing atomic hydrogen to be displaced and only partially removed \citep{scott_2010}, while the denser, more centrally located molecular gas is found to be essentially unperturbed \citep{boselli_1997, vollmer_2001}.  
By contrast, in the most massive clusters the environment encountered by infalling galaxies can lead to their entire gas reservoir being stripped in a single pass through the cluster core \citep[e.g.][]{takeda_1984, abadi_1999, kapferer_2009, steinhauser_2012}. 
Observational evidence of extreme ram-pressure stripping is, however, sparse, due to 
their reliance on favourable circumstances, such as suitable infall trajectory, gas mass, galaxy orientation, and high ICM density. Considering the small number and relatively low masses of nearby clusters (except for Coma), these conditions are unlikely to be met in the local Universe.

The extreme environment that is a prerequisite for extreme RPS is, however, routinely encountered by galaxies falling into massive clusters where galaxy peculiar velocities in excess of 1000 km s\textsuperscript{-1} are common and the ICM particle density easily exceeds 10\textsuperscript{-3} cm\textsuperscript{-3}. Since massive clusters are rare, larger volumes have to be searched to efficiently probe such truly high-density environments.  Although their numbers are still small, striking examples of extreme RPS events have been discovered in \emph{Hubble Space Telescope} (\emph{HST}) images of moderately distant ($z\gtrsim0.2$) massive clusters \citep{owen_2006, cortese_2007, owers_2012} and, most recently, in X-ray selected massive clusters at $z{>}0.3$ \citep[][see Fig.~1]{ESE}.  Importantly, these clusters are not only intrinsically more massive, they are also dynamically less evolved and more likely to be undergoing mergers than systems in the local Universe \citep{mann_2012}, a critical requirement if extreme RPS events are triggered by merger-driven shocks, as suggested by \citet{owers_2012}. Increasing the size of the still small sample of RPS examples clearly constitutes a crucial step toward a meaningful statistical investigation of the physics of accelerated galaxy evolution.

In this paper, we aim to compile a statistically significant sample of galaxies that might be undergoing RPS in very massive clusters. We then use this sample to establish which galaxy trajectories are most conducive to creating extreme RPS, and thereby elucidate whether the most dramatic RPS events are triggered by massive cluster mergers \citep{owers_2012}, rather than during regular infall of galaxies from the field or along filaments.  In order to compile the required sample, we develop morphological criteria to select RPS candidates from archival \emph{HST} imaging data for a well defined sample of massive clusters at $z>0.3$, and compare the spatial and dynamical distribution of the selected objects with expectations from numerical and theoretical models.

This paper is structured as follows: 
in Section~\ref{sec:data} we introduce the cluster sample and present an overview of the observations and data-reduction procedures; 
in Section~\ref{sec:morph} we discuss our morphological criteria for the identification of galaxies experiencing ram-pressure stripping and present the sample of RPS candidates;
in Section~\ref{sec:models} we present the a simple model of clustre infall which we use to interpret our data;
in Section~5 we present our results for the spatial distribution and dynamical properties of RPS events in massive clusters; and in Section~6 we draw conclusions about the origin, trajectories, and physics of extreme RPS. We present a summary of our work in Section~7.

Throughout this paper, we assume a concordance $\Lambda$CDM cosmology with $\Omega_M$ = 0.3, $\Omega_\Lambda$ = 0.7, H$_0$ = 70 km s\textsuperscript{-1} Mpc\textsuperscript{-1}.
As the clusters in our sample span a range of redshifts of $0.3<z<0.7$, the metric scale of our images varies from 4.45 to 7.15 kpc arcsec\textsuperscript{-1}. 

\begin{figure*}
 \centerline{
  \includegraphics[width=0.3\textwidth]{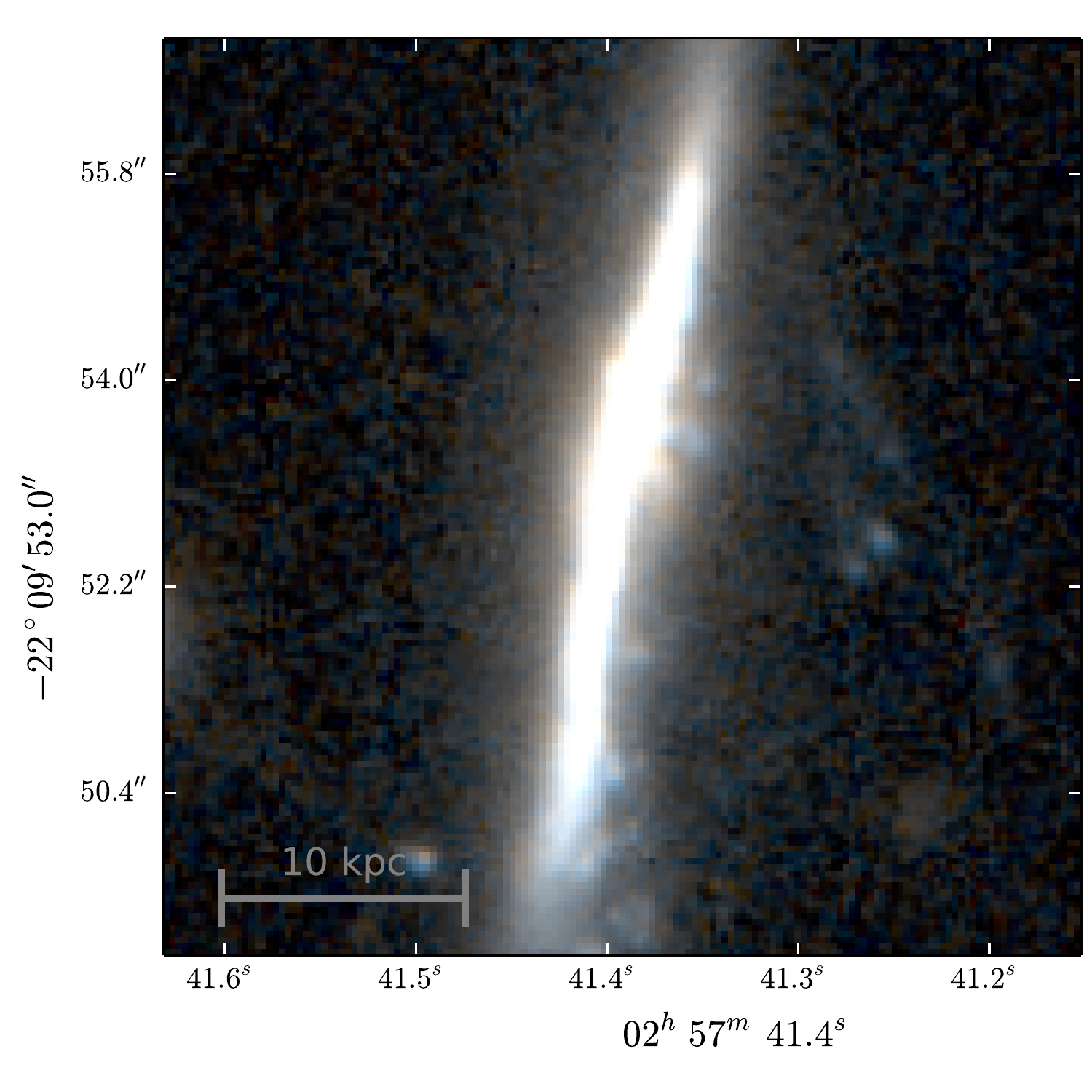}
  \includegraphics[width=0.3\textwidth]{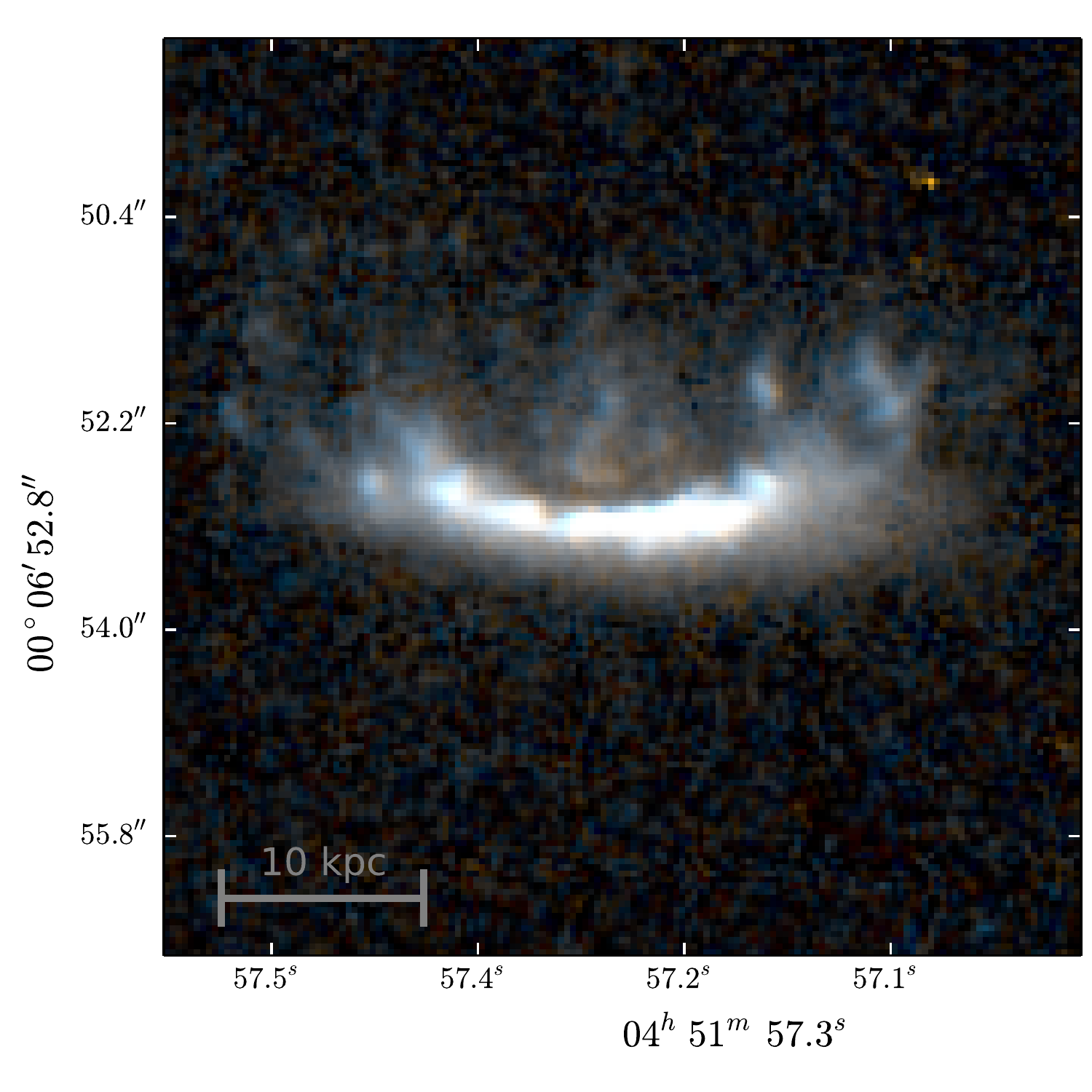}
  \includegraphics[width=0.3\textwidth]{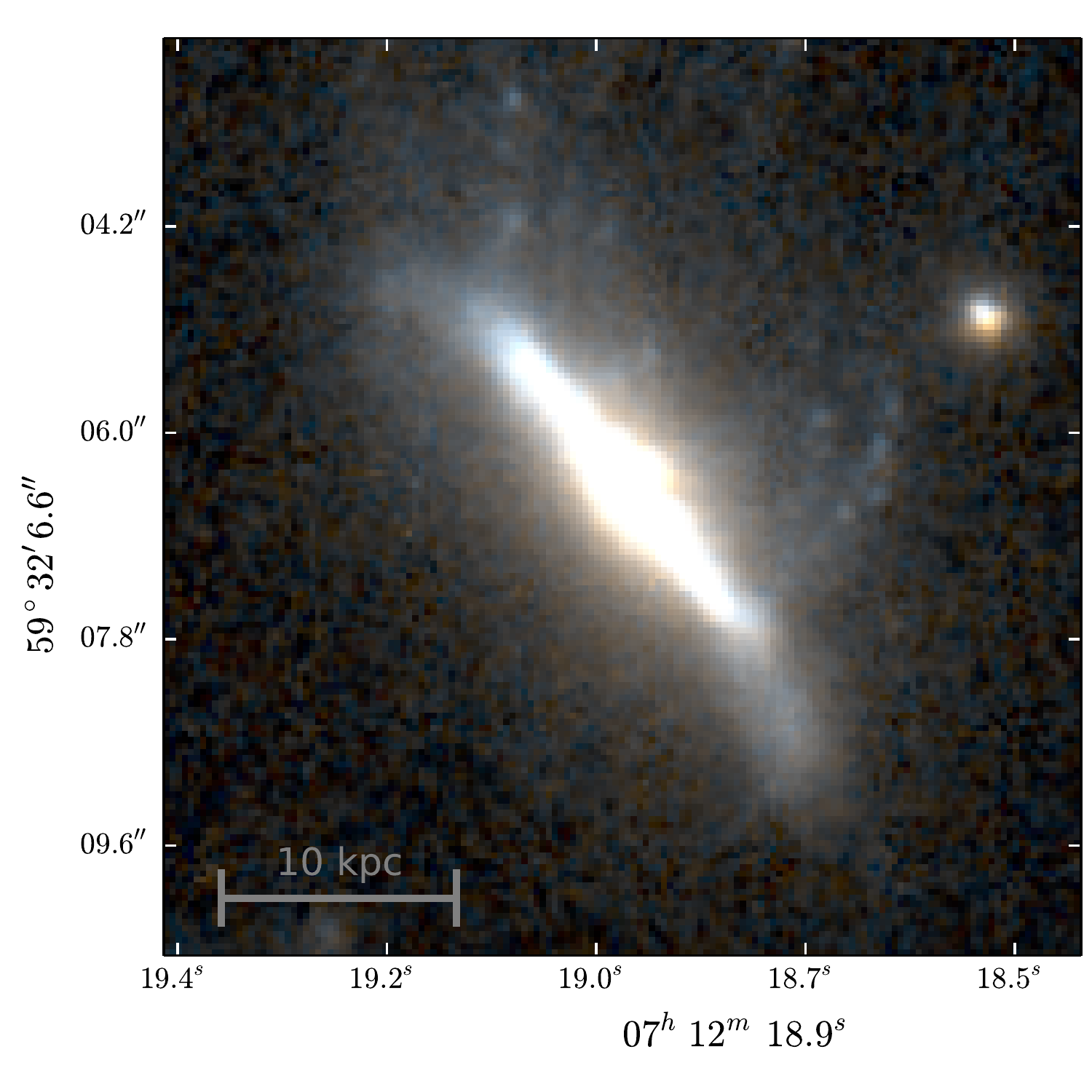}
 }
 \centerline{
  \includegraphics[width=0.3\textwidth]{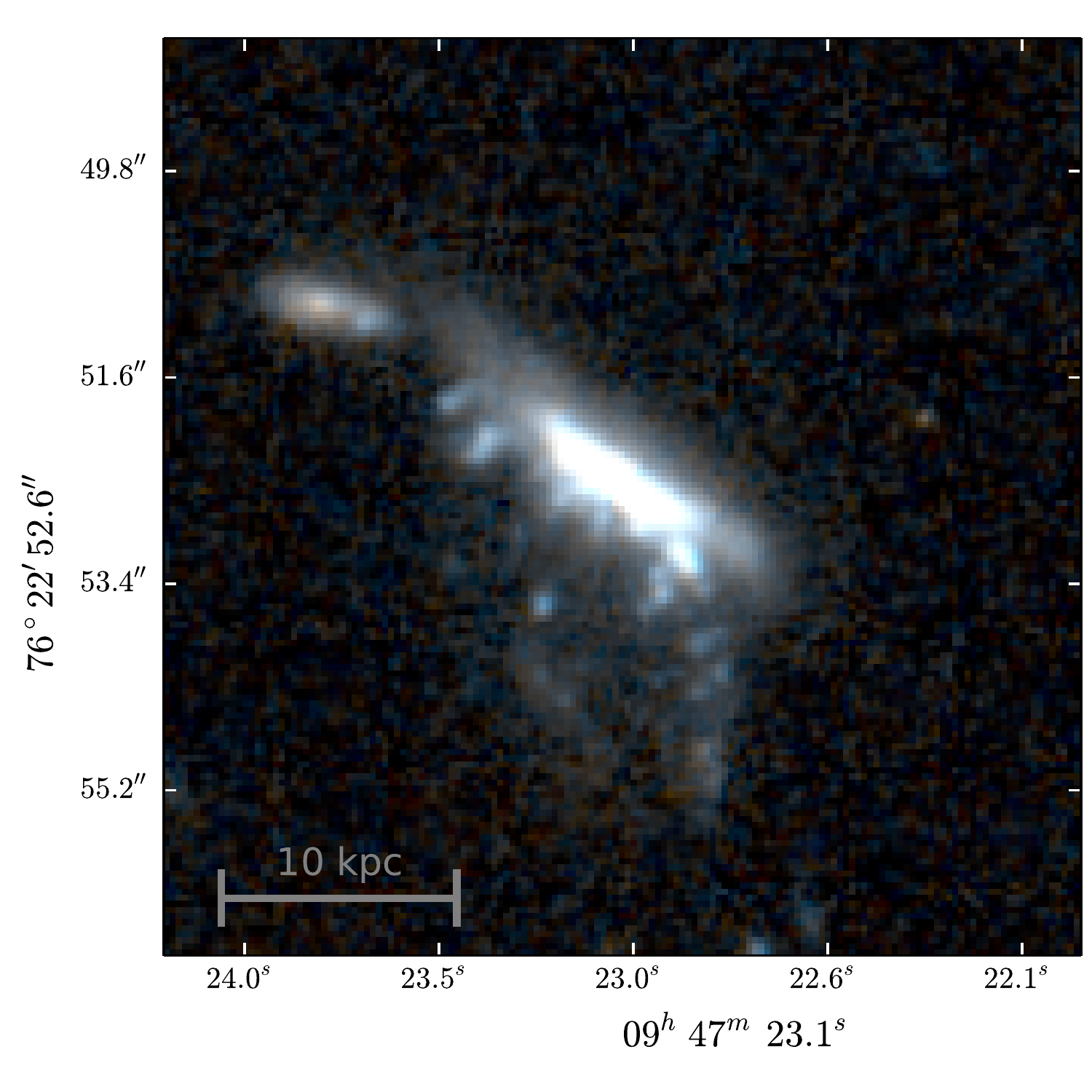}
  \includegraphics[width=0.3\textwidth]{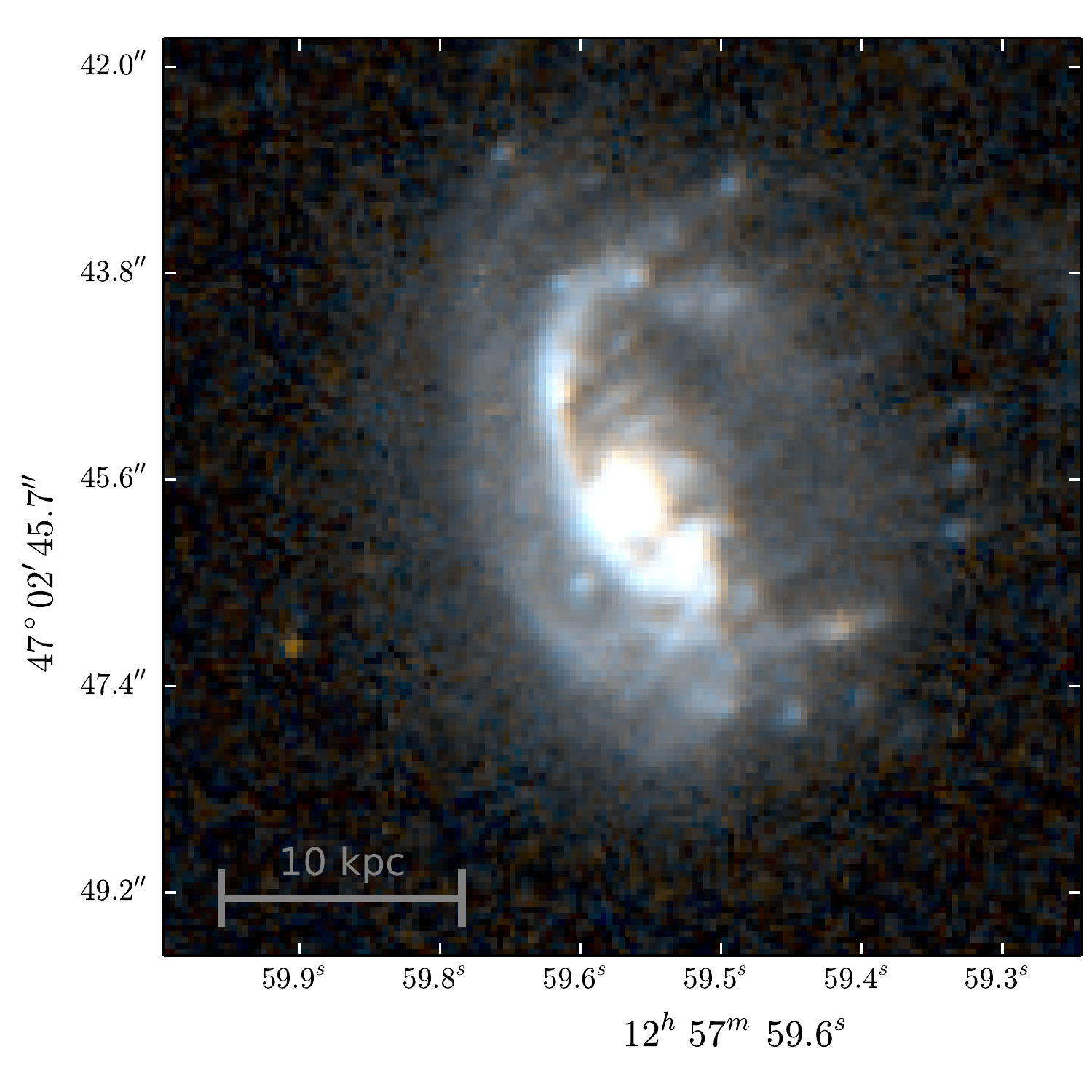}
  \includegraphics[width=0.3\textwidth]{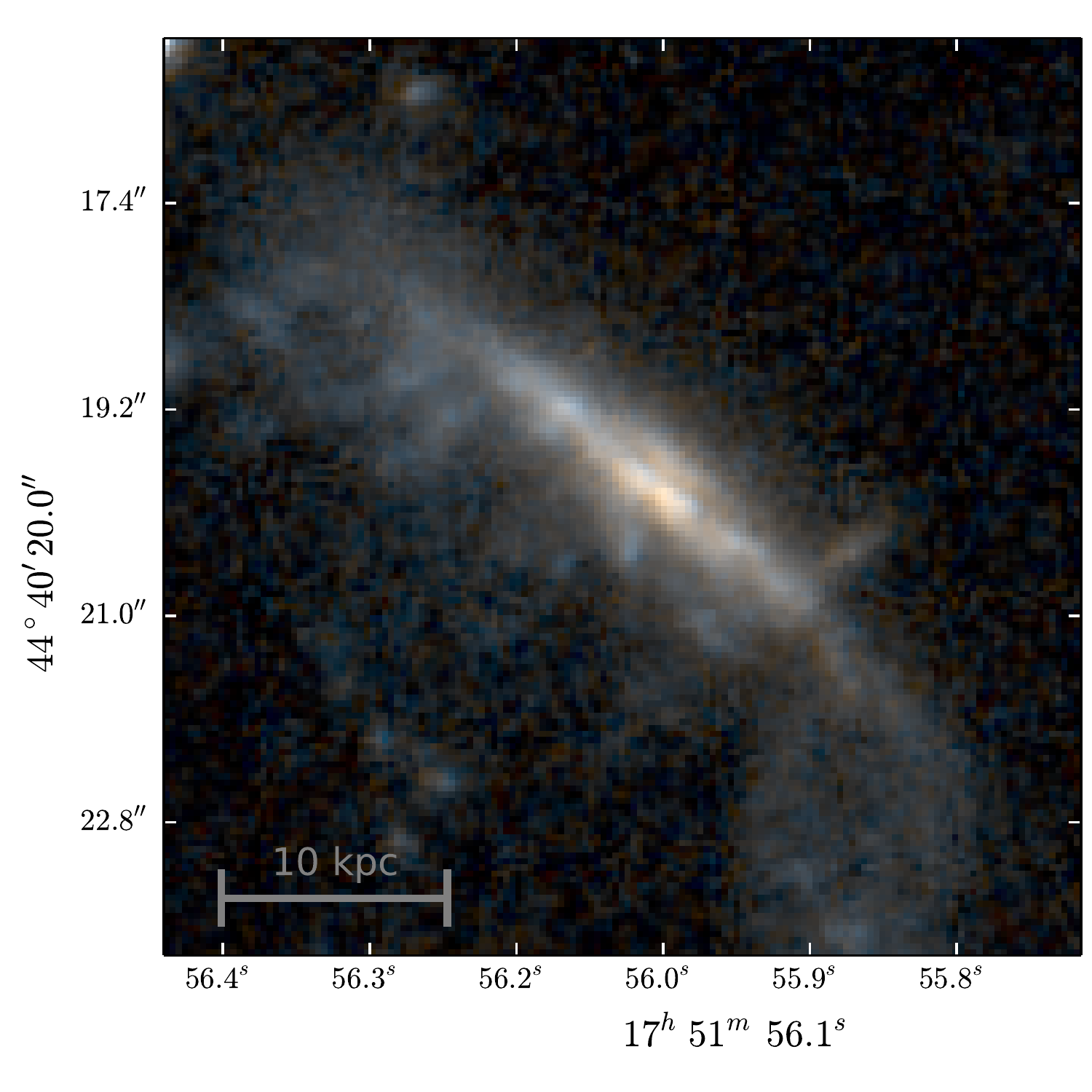}
 }
 \centerline{
  \includegraphics[width=0.3\textwidth]{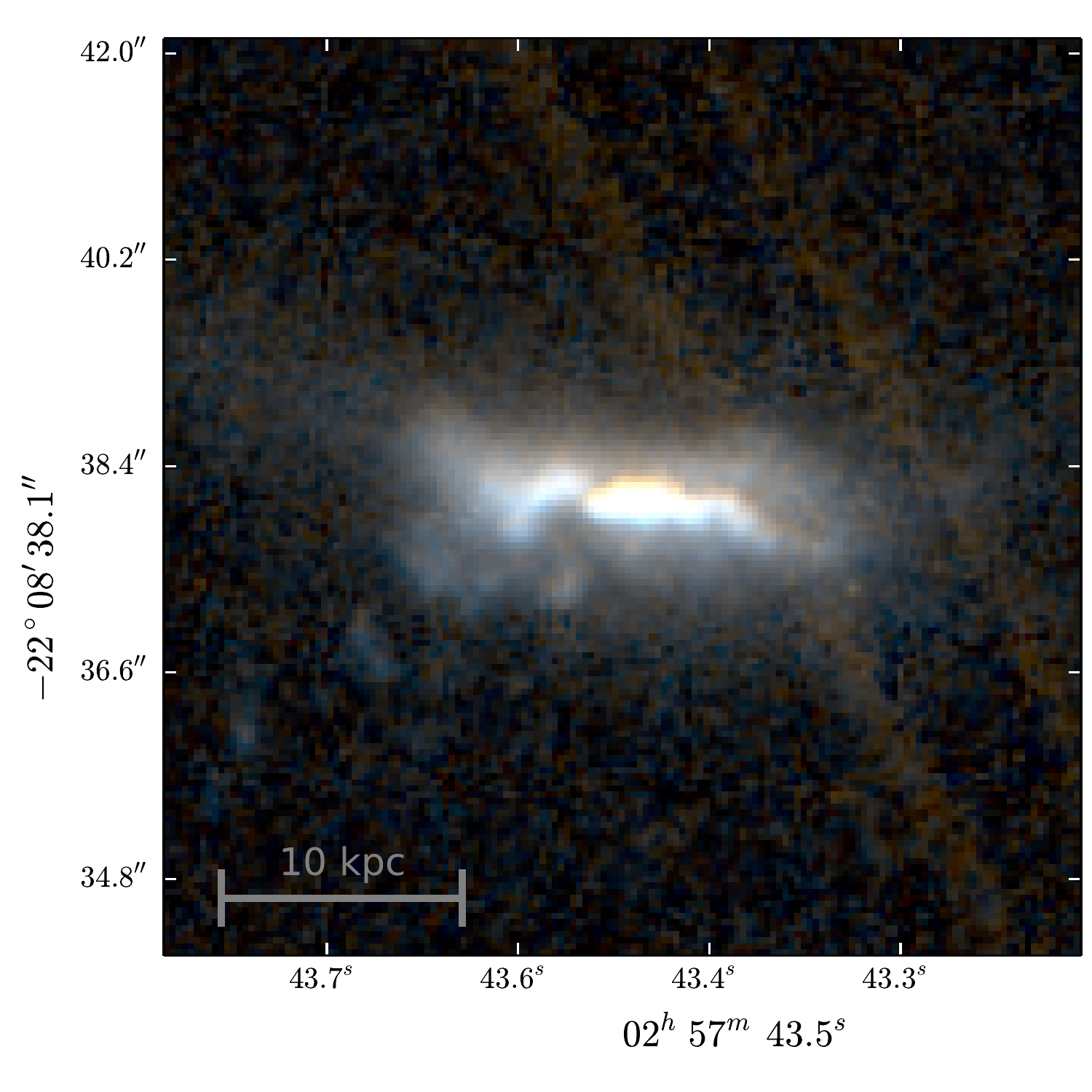}
  \includegraphics[width=0.3\textwidth]{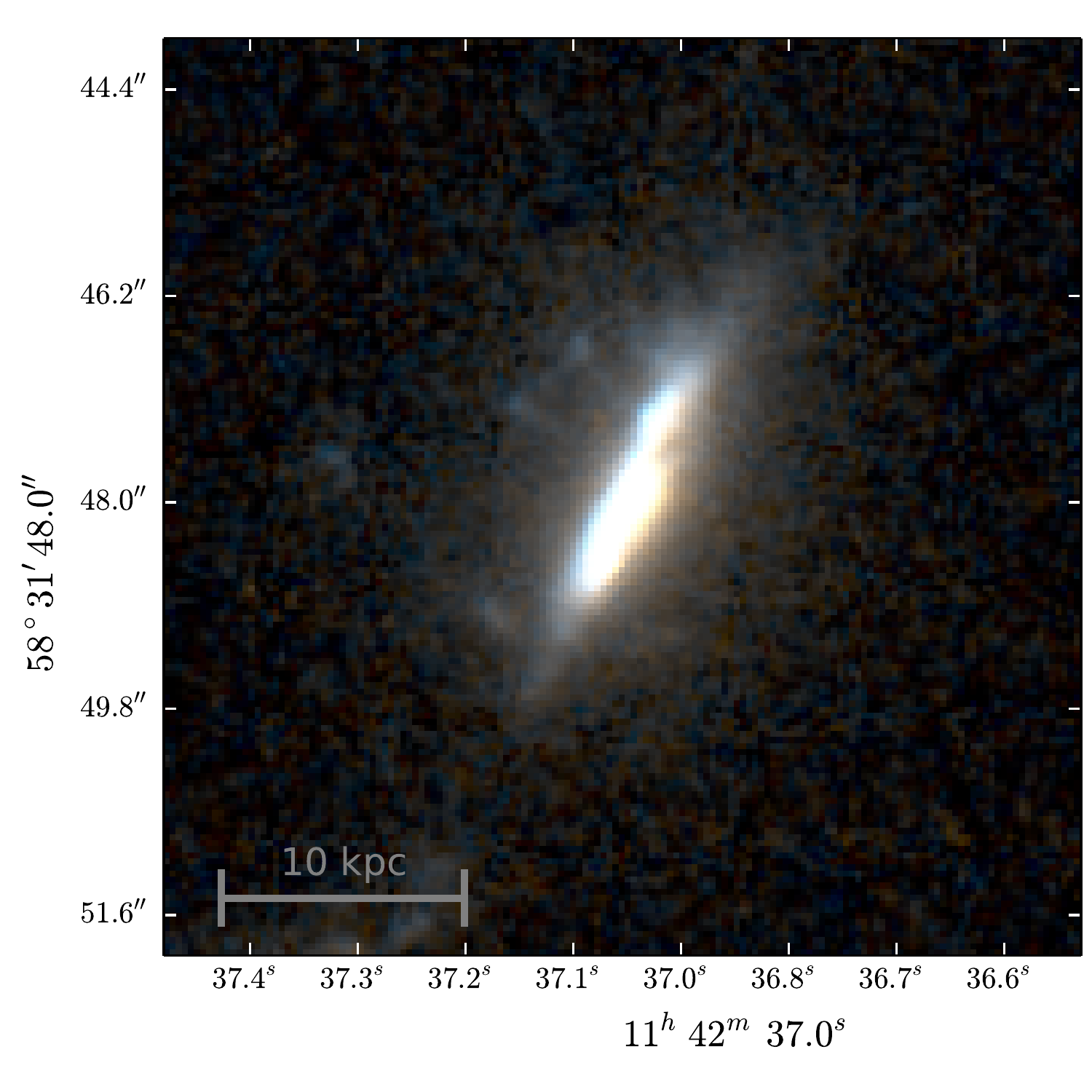}
  \includegraphics[width=0.3\textwidth]{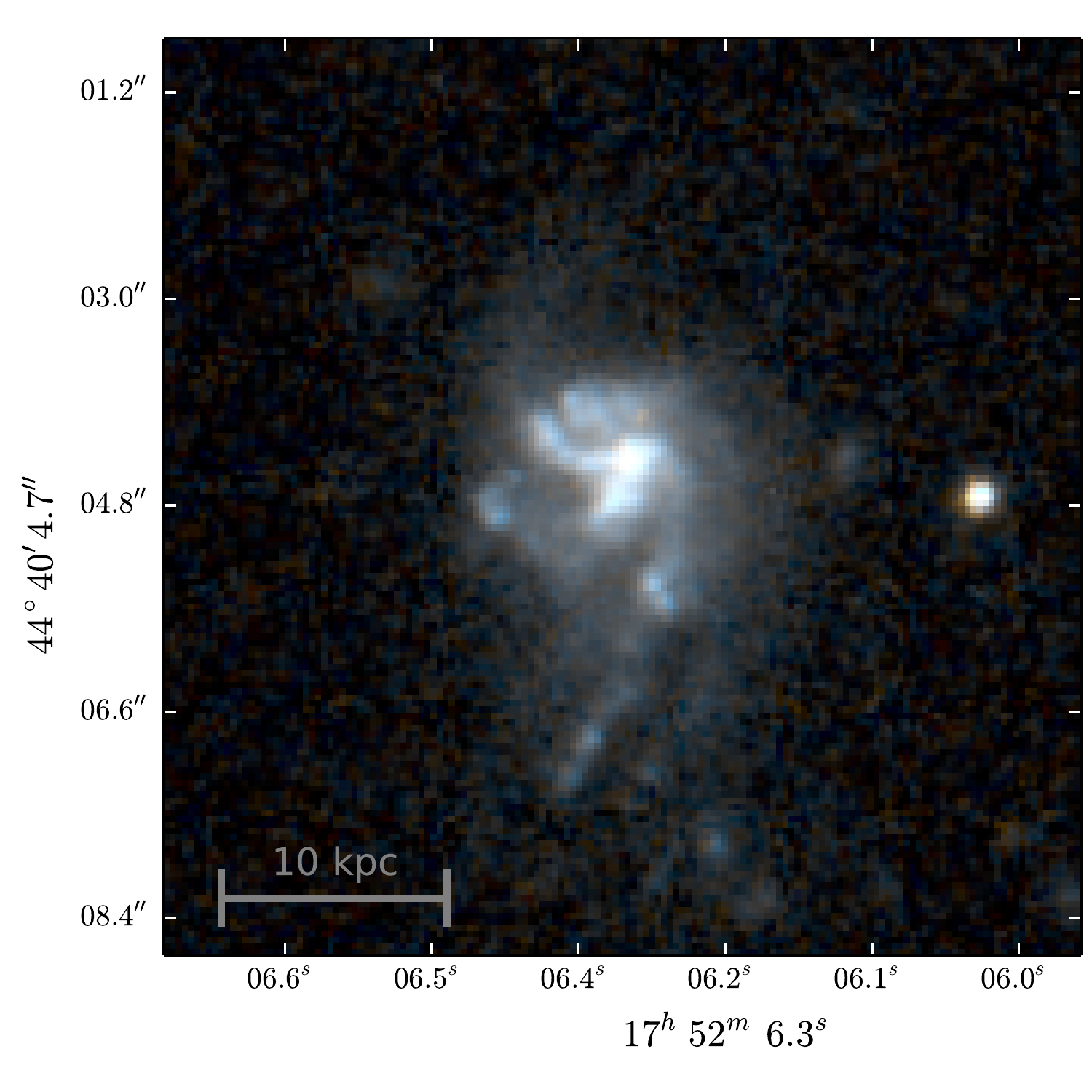}
 }
 \begin{mdframed}[innerleftmargin=5pt,innerrightmargin=5pt, linecolor=red]
  \centerline{
   \includegraphics[width=0.3\textwidth]{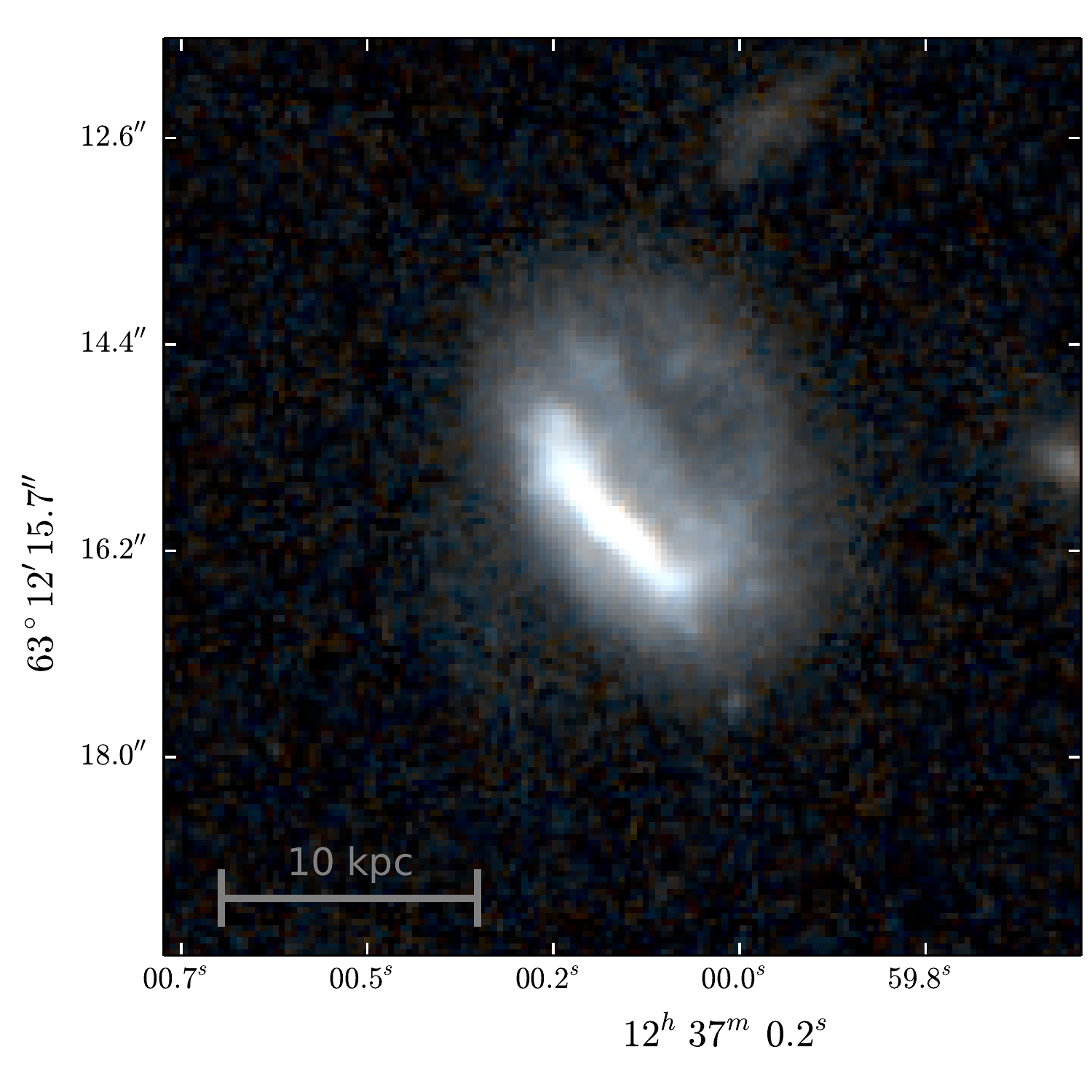}
   \includegraphics[width=0.3\textwidth]{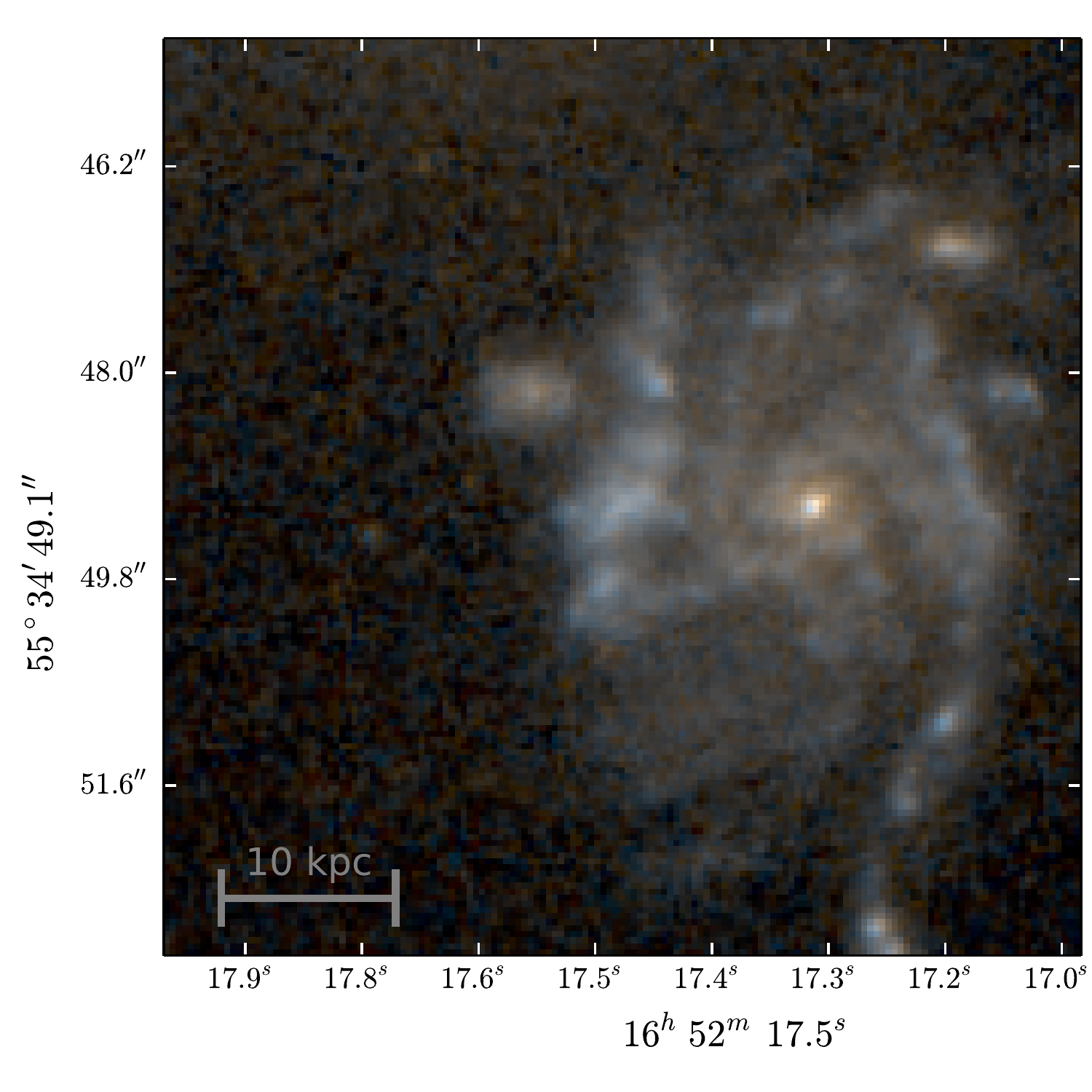}
   \includegraphics[width=0.3\textwidth]{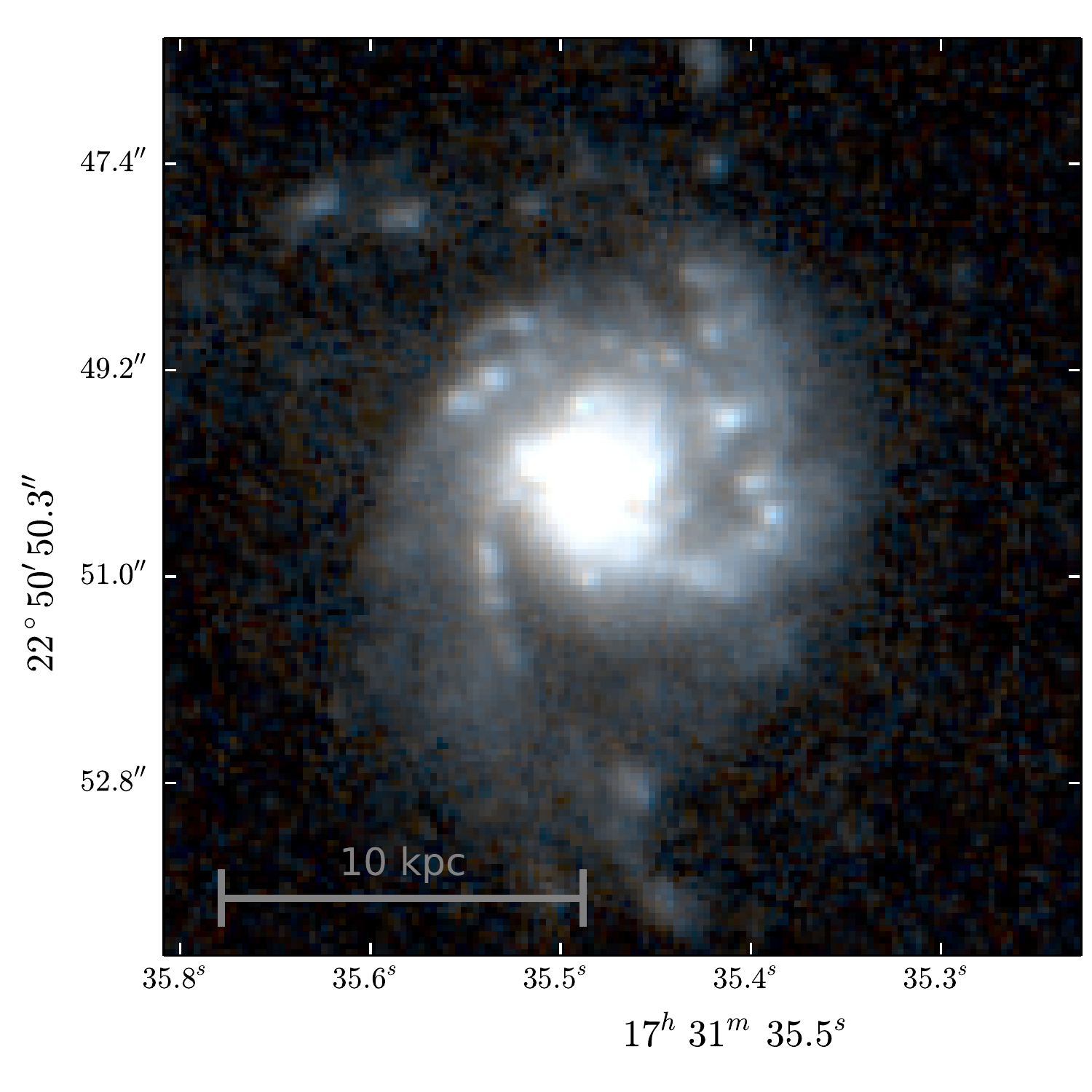}
  }
 \end{mdframed}
 \caption{The 12 galaxies deemed textbook examples of ram-pressure stripping and thus used as our training set; six of these (top two rows) were published previously by \citet{ESE}. Three members of our training set were recently found not to be cluster members (see Section~\ref{sec:bias}) and are highlighted in the bottom row.\label{fig:ese}}
\end{figure*}

\begin{figure*}
\centerline{
\hspace*{-0mm} \includegraphics[width=0.35\textwidth]{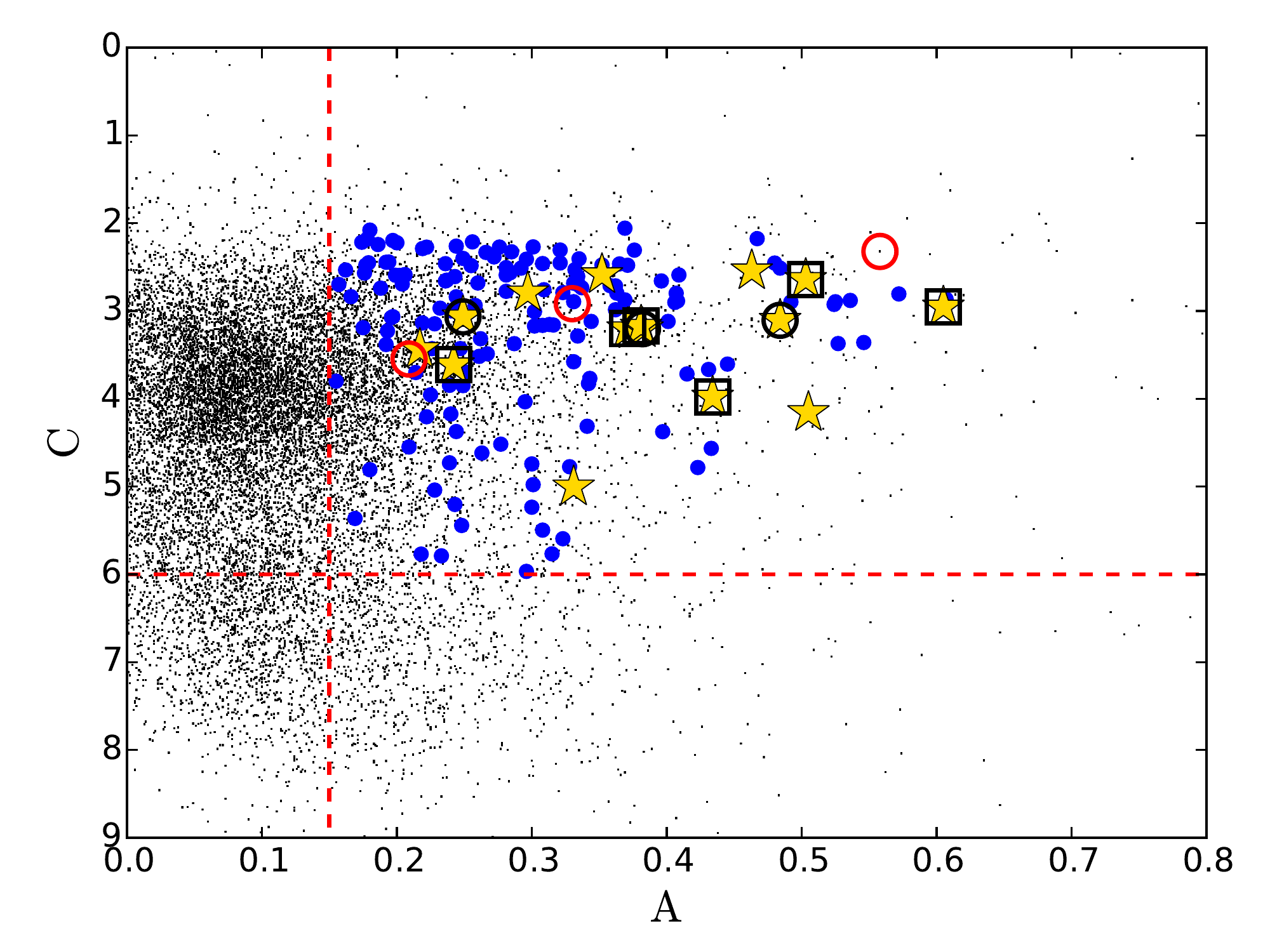}
\hspace*{-3mm}  \includegraphics[width=0.35\textwidth]{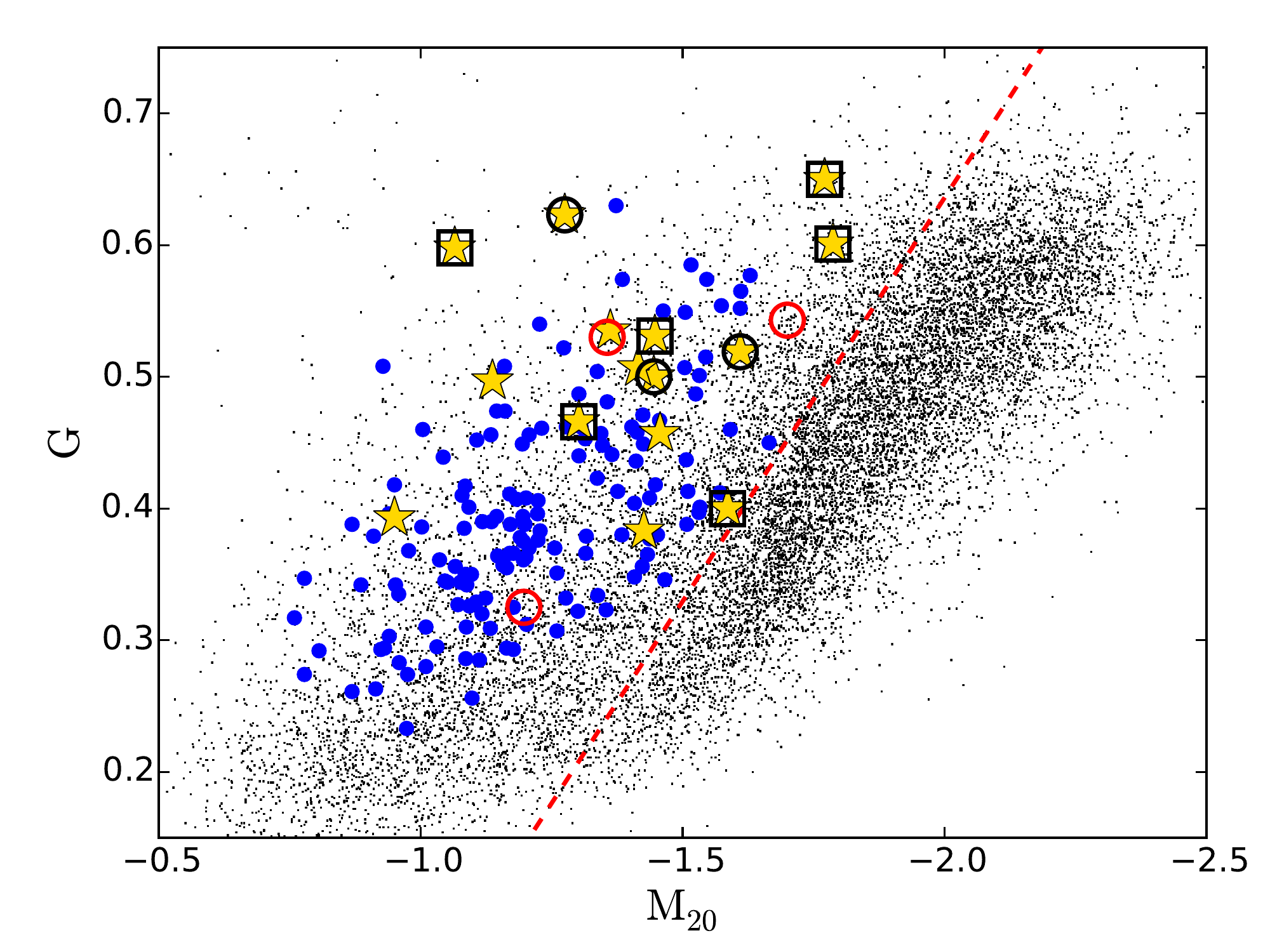}
\hspace*{-3mm}  \includegraphics[width=0.35\textwidth]{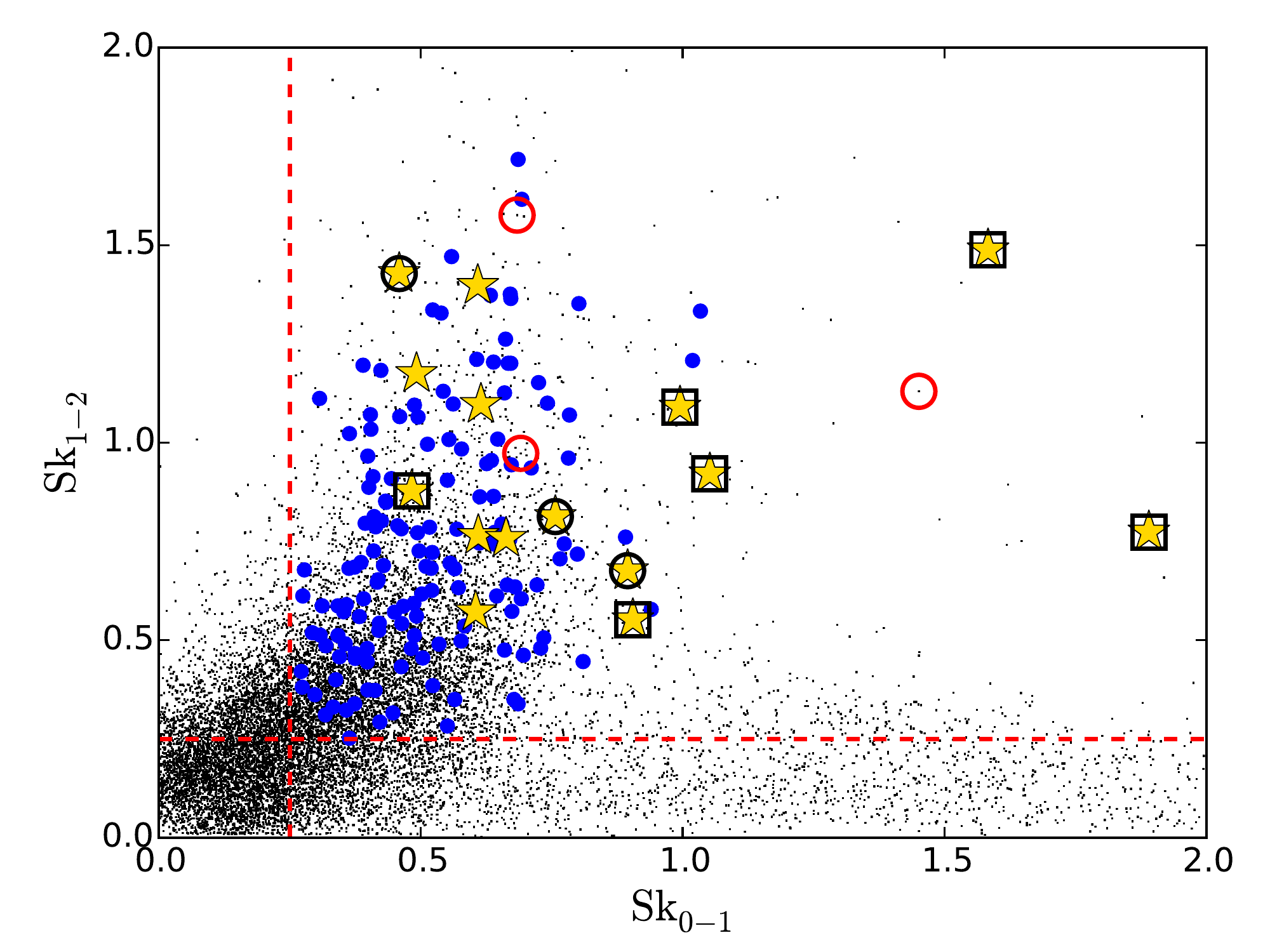}
}
\caption{The distribution of all galaxies in our target fields in various parameter spaces. Left: Concentration--Asymmetry; centre: Gini--$M_{20}$; right: $Sk_{0-1}$--$Sk_{1-2}$. Our final sample of RPS candidates is marked by filled blue circles; the morphologically most compelling examples are shown as yellow asterisks.  Members of our training set (see Fig.~\ref{fig:ese}) are shown with open symbols. Squares show the six systems published in \citet{ESE}, and circles show the six additional galaxies from their extended sample. Three members of our training set, all part of the extended ESE sample, were recently found not to be cluster members (see Section~\ref{sec:bias}) and are shown in red. The cuts defining our final morphological selection criteria are indicated by red dashed lines.  \label{fig:cuts}}
\end{figure*}

\section{Data used in this study}
\label{sec:data}
\subsection{The MACS sample}
Our cluster sample is drawn from a master list of clusters identified in the course of the Massive Cluster Survey \citep[MACS;][]{ebeling_2001,ebeling_2007,ebeling_2010,mann_2012}, designed to provide a large, statistically complete sample of X-ray luminous ($L_X \gtrsim 5\times 10^{44}$ erg s\textsuperscript{-1}, 0.1-2.4 keV) and moderately distant ($z \gtrsim 0.3$) galaxy clusters. Covering over 22,000 sq.deg., the MACS sample comprises the majority of massive galaxy clusters in the observable Universe, making it ideally suited for our investigation. At redshifts $z\gtrsim0.3$, the sub-kiloparsec angular resolution needed to identify the characteristic morphological traits of RPS events can only be achieved with the Advanced Camera for Surveys (ACS) aboard \emph{HST}.  We thus limit our sample to MACS clusters with archival \emph{HST}/ACS images as described in more detail in the following section.

\subsection{Imaging data}
 As our primary observational diagnostics revolve around morphological features traced by star-forming regions, we limit our study to MACS clusters that have been observed in the \emph{HST}/ACS F606W band.  The F606W filter is well suited as it corresponds roughly to the B band in the cluster rest frame and has been used in a large number of \emph{HST} observations of MACS clusters. We further require clusters in our sample to also have imaging data in the ACS F814W passband, as the resulting F814W--F606W colours provide a straightforward means to discriminate against the population of passively evolving cluster ellipticals.

Of the entire MACS sample, 44 clusters were successfully observed in both the ACS F606W and F814W passbands as part of the \emph{HST} SNAPshot programmes GO-10491, -10875, -12166, and -12884 (PI: Ebeling). These programmes use short exposures (1200 seconds for F606W and 1440 seconds for F814W) designed to reveal bright strong-lensing features and provide constraints on the physical nature of galaxy-galaxy and galaxy-gas interactions in cluster cores.  Fundamental properties of this subset of the MACS cluster sample are presented and discussed by Ebeling \& Repp (in preparation). Supplementing these SNAPshots, we also include data from observations of 17 additional MACS clusters obtained by the Cluster Lensing and Supernova Survey with Hubble \citep[CLASH;][]{CLASH}, an \emph{HST} Multi-Cycle Treasury Program employing 16 filters from the UV to the NIR, including F606W and F814W. Exposure times for the CLASH observations are nominally one and two orbits for all ACS filters, but vary substantially between cluster fields around median exposure times of 4060 and 8480 seconds for the F606W and F814W passbands, respectively (see Table \ref{tab:snaps_obs} \& \ref{tab:clash_obs} for a summary of the observations).

In total, our sample thus comprises 63 MACS clusters. At the redshifts relevant to our study, the field of view of the ACS Wide Field Channel ($202^{\prime\prime}\times202^{\prime\prime}$) covers an inscribed circle of radius between 450 and 720 kpc and thus samples primarily the cluster core region. Charge-transfer-efficiency corrected images in the two passbands were registered using the astrometric solution of the F606W image as a reference, and source catalogs were created using SExtractor \citep{sextractor} in dual-image mode, with F606W chosen as the detection band. We removed stars as well as cosmic rays and other artefacts as objects falling on or below the star lines in both magnitude-$\mu_\mathrm{max}$ and magnitude-$r_{20\%, \mathrm{ell}}$ space\footnote{Here, $\mu_\mathrm{max}$ and $r_{20\%, \mathrm{ell}}$ are the peak surface brightness and the elliptical radius encircling 20\% of the total flux, respectively.}. After removing spurious detections, we have a 5$\sigma$ 90\% completeness limiting magnitude of 24.9 in F606W (here and in the following the magnitudes quoted are measured within the Petrosian radius). 

As the quantitative morphological indicators we employ to identify RPS candidates (see Section~\ref{sec:morph}) require signal-to-noise  ratios of $\langle$S/N$\rangle{>}5$ per pixel, we limit our galaxy sample to objects with $m_{\rm F606W}{<}24$, which leaves a total of 15,875 galaxies (11,550 in the SNAPshot data and 4,325 in the CLASH data). We note that, due to the high density of objects in cluster cores and the presence of objects of complex morphology, some of the objects in our master catalogue may in fact be blends of several objects, whereas others have suffered fragmentation, i.e., were broken up into multiple sources. 

To mitigate the effect of fragmentation in our master catalog, we enforce strict deblending criteria (DEBLEND\_NTHRESH=16, DEBLEND\_MINCONT=0.2). Due to the relatively shallow depth ($\sim$1200 s) of the imaging data, the faint extraplanar tails that characterize jellyfish galaxies often fall below our detection limit. For the quantitative selection criteria (see Section~3.1), we, therefore, focus on identifying robust morphological features (disturbances) in the high signal to noise regions of galaxies. However, note that the presence of optical tails is a requirement for an object to be classified as a compelling jellyfish candidate during our visual screeening process. 

As for the completeness of the sample of candidates presented here, it is almost certain that modest cases of RPS (in particular when occuring in low mass galaxies) will have been missed due the lack of pronounced morphological features, whereas essentially all the brightest objects would have been easily identified by eye. We note however that regarldless of brighness, objects moving close to our line of sight are likely to be missed as the tell-tale debris trails will be obscured by the the much brighter disks of the galaxies. We discuss this bias in detail in Section~\ref{sec:bias} and \ref{sec:bias2}.

\subsection{Spectroscopic data}

The sample of RPS candidates compiled in this work using morphological selection is expected to be heavily contaminated by galaxies that are in fact not members of the respective MACS cluster and / or whose morphology is irregular for reasons other than RPS (see Section~\ref{sec:morph} for details).  In order to eliminate interlopers, we have embarked on a comprehensive spectroscopic survey of our RPS candidate sample, aimed at (a) excluding fore- and background galaxies from our sample of RPS candidates, and (b) obtaining peculiar radial velocities of those systems that are cluster members.  We refer to a forthcoming paper (Blumenthal et al., in preparation) for a more extensive report on these efforts, including a description of the data-reduction procedure.  We note here though that all spectroscopic observations were conducted with the DEIMOS spectrograph on the Keck-II 10m-telescope on Maunakea, using multi-object spectroscopy with slits of 1mm width, the 600 l/mm Zerodur grating set to a central wavelength of 6300\AA, the GG455 blocking filter, and exposure times ranging from 3$\times$10 to 3$\times$15 minutes.  For almost all targeted galaxies, redshifts were measured from emission lines detected in these spectra, yielding a precision of approximately 0.0002 in redshift or 60 km s$^{-1}$ in radial velocity.

\section{GALAXY MORPHOLOGY}
\label{sec:morph}

A recent study by \citet[][hereafter ESE]{ESE} presented six textbook examples of ``jellyfish" galaxies (thought to be extreme RPS events\footnote{Although the observed morphology of these objects does not prove the occurrence of RPS, in-depth follow-up studies of galaxies sharing the same striking features unambiguously confirmed RPS to be at work (\citealt{sun_2010}; \citealt{sivanandam_2010}; \citealt{cortes_2015}).}) discovered in \emph{HST} imaging data for 36 of the 63 clusters used in this work. These objects were visually identified, having to meet the following criteria: (1) a strongly disturbed morphology in optical images indicative of unilateral external forces; (2) a pronounced brightness and colour gradient suggesting extensive triggered star formation; (3) compelling evidence of a debris trail. Furthermore, the direction of motion implied by each of these features had to be consistent. We expand the ESE sample by six additional, unpublished, jellyfish candidates, identified by the same authors, that satisfy at least two of these criteria\footnote{Note that the inferred direction of motion for two candidates (leftmost two in the bottom row of Fig.~\ref{fig:ese}) is largely aligned with our line of sight.}, and use the resulting superset of 12 objects (shown in  Fig~\ref{fig:ese}) as a training set for the identification of additional, less obvious candidate objects.

For each of the galaxies in our catalogue we compute several non-parametric galaxy morphology statistics defined previously in the literature: concentration ($C$) and  asymmetry ($A$) \citep{bershady_2000,CAS}, Gini coefficient ($G$) and $M_{20}$ \citep{abraham_2003, lotz_2004}. While these statistics were originally designed to identify the morphological features of galaxy mergers, we find that they can be applied more widely to characterise and select objects featuring disturbed morphologies.  In addition to the aforementioned four statistics, we introduce two ``skeletal decomposition" parameters ($Sk_{0-1}$ and $Sk_{1-2}$; see Appendix~\ref{sec:skel}). 

We compute values for each of these indicators using the ellipticities, position angles, and locations provided by SExtractor but note that the precise location of the centre of each object is iteratively refined through minimisation procedures, as described in \citet{lotz_2004}.  Acknowledging the difficulty of cleanly separating galaxies in crowded cluster cores, we resort to using SExtractor's segmentation maps to identify the pixels belonging to a given galaxy rather than relying on an isophotal definition of a galaxy's extent.  We stress that, as a result, the morphological quantities measured here should not be directly compared to those from other work.

\begin{figure}
 \centerline{
\hspace*{-0mm} \includegraphics[width=0.3\textwidth]{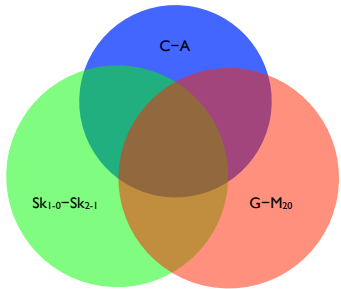}
 }
 \caption{Venn diagram of the sets of galaxies selected by each of the morphological criteria shown in the three panels of Fig.~\ref{fig:cuts}. Although each type of cut selects a similar number of galaxies (represented by the area of each circle), the modest overlap between these sets makes the final selection, achieved by requiring all criteria to be met, much more restrictive. \label{fig:venn}}
 \end{figure}

 \subsection{Selection criteria and visual screening}
 
The fact that the extended ESE sample (Fig.~\ref{fig:ese}) contains some of the most extreme examples of jellyfish galaxies known to date (i.e., the brightest and most morphologically disturbed) makes it well suited as a training set for an iterative, semi-automated search for additional RPS candidates. To this end, we examine the location of the training-set members in $C$--$A$, Gini--$M_{\rm 20}$, and $Sk_{0-1}$--$Sk_{1-2}$ space, and define cuts in these parameter spaces that preserve the training set but eliminate the vast majority of other galaxies.  The physical rationale behind these cuts is to discard extremely diffuse objects (achieved by a cut in $C$), almost perfectly symmetric sources (cut in $A$), morphologically undisturbed disk and elliptical galaxies (cut in $G$-$M_{20}$), and, finally, objects with little substructure (cuts in $Sk_{0-1}$ and $Sk_{1-2}$).

 
We apply an initial set of morphological criteria (cuts in $C$--$A$ and Gini--$M_{20}$) to galaxies detected in the 10 cluster fields from which the extended ESE sample originates.  The $\sim$650 candidate objects thus selected are then visually scrutinised independently by two of us (CM and HE) and classified according to their plausibility as RPS events.  We attempt to reduce the subjectivity of this procedure by reviewing jointly, in a second iteration, all objects classified either as compelling jellyfish galaxies or as plausible candidates by one of the inspectors and assigning a consensus classification. From the resulting set of potential RPS events we select the most compelling candidates, add them to our original training set, and re-evaluate our initial morphological constraints. Cuts in colour-magnitude space were considered too during this process but ultimately dismissed as largely redundant with the aforementioned morphological cuts, which already remove the majority cluster ellipticals and faint blue objects. The full set of morphological criteria (now also including cuts in $Sk_{0-1}$--$Sk_{1-2}$) are then applied to the remaining clusters, and the resulting subset is once again visually screened.  Fig.~\ref{fig:cuts} shows the distribution of all galaxies in various projections of our multi-dimensional morphology parameter space, as well as the applied selection criteria.  Members of the extended training set and of our final sample of RPS candidates are highlighted.  Although the three sets of selection criteria shown in Fig.~\ref{fig:cuts} all select approximately the same fraction of galaxies (30-40\%), their doing so largely non-redundantly leads to a much more restrictive selection of merely 8\% (1263 galaxies) when all criteria are combined (Fig.~\ref{fig:venn}).

\begin{figure}
  \includegraphics[width=0.24\textwidth]{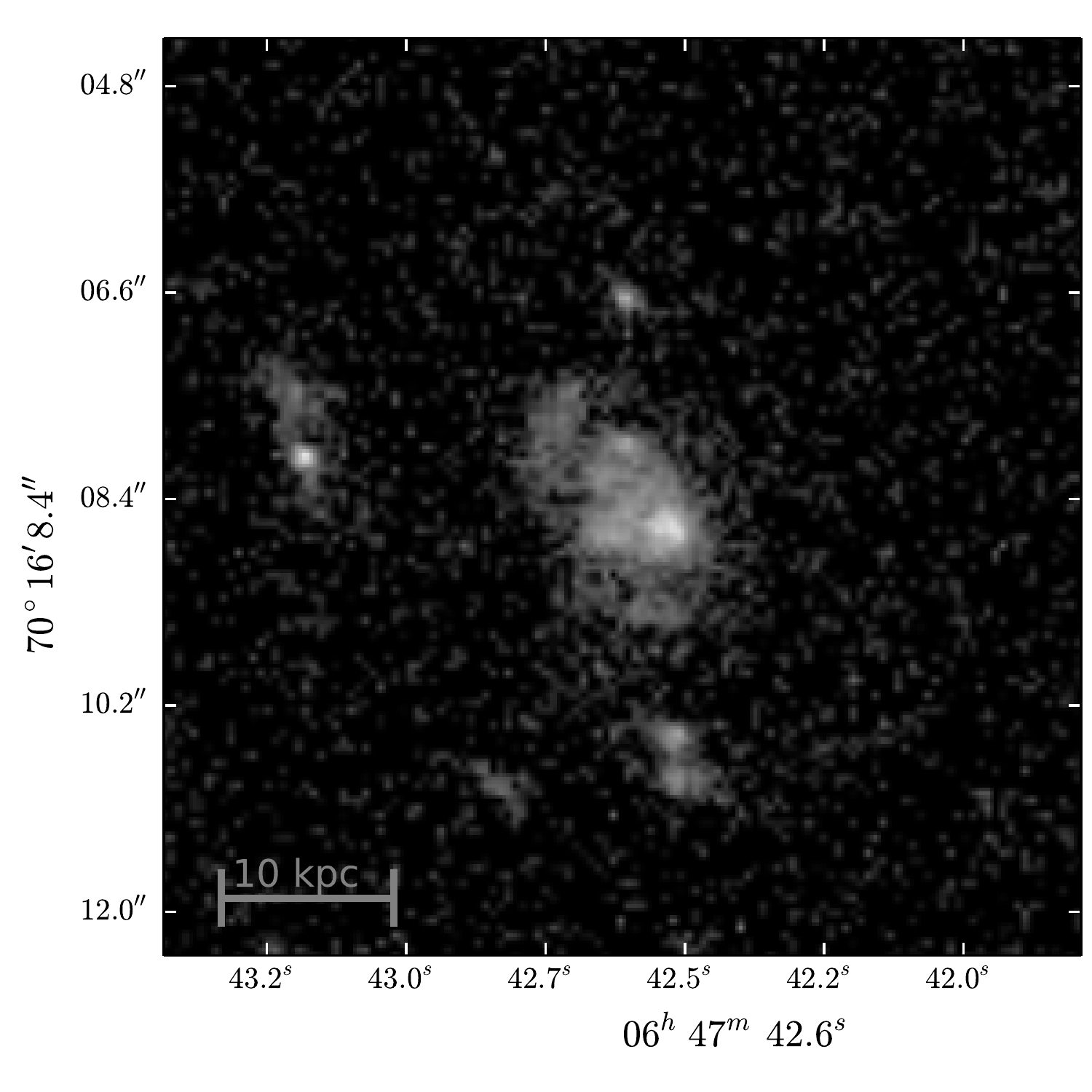}
  \includegraphics[width=0.24\textwidth]{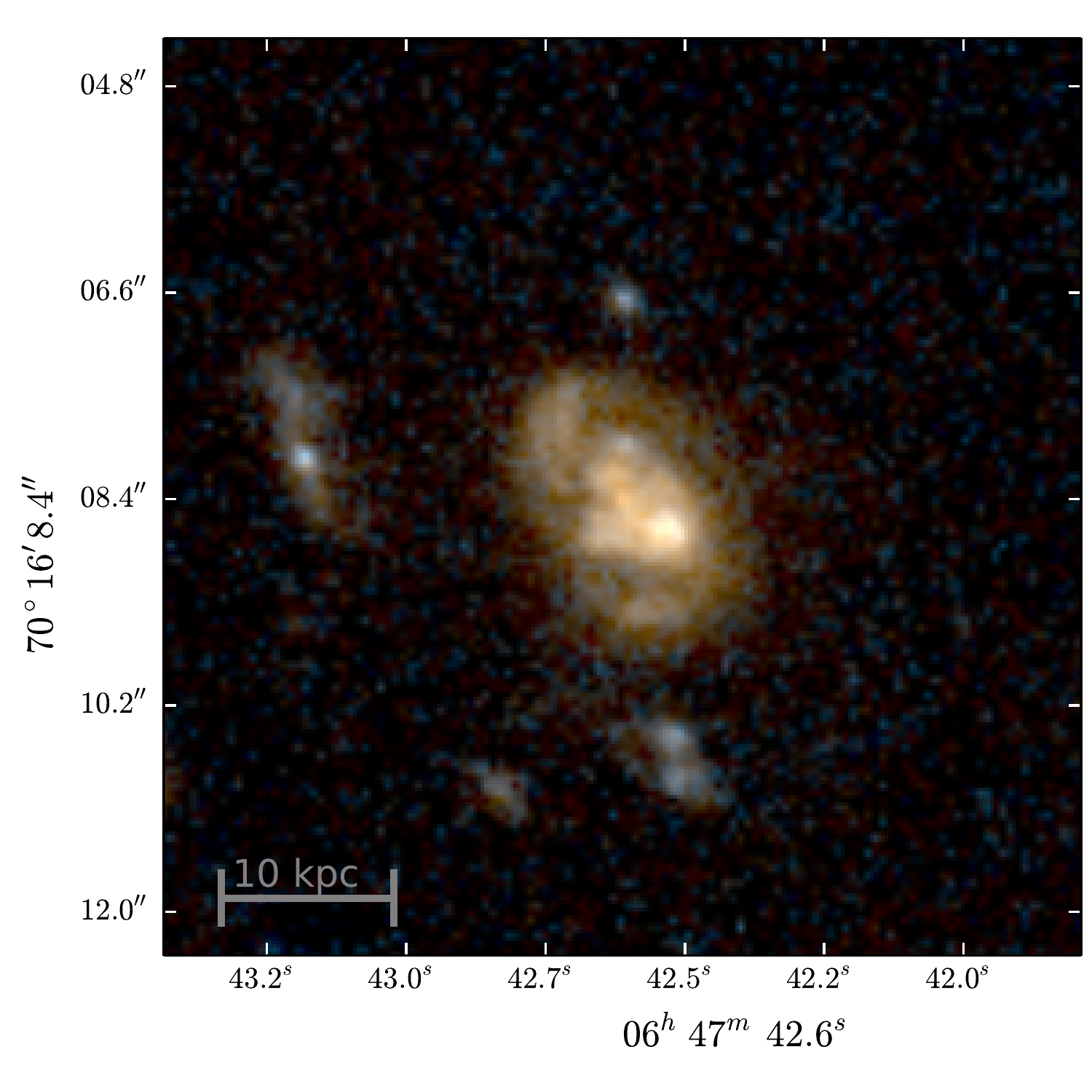}
  \caption{Importance of colour information for our visual inspections. Viewed solely in the F606W passband (left) this object could be considered a (remotely) plausible RPS candidate.  A false-colour image including data in the F814W filter (right) strongly suggests a slightly disturbed dusty disk galaxy. \label{fig:need_color}}
 \end{figure}

It is evident from Fig.~\ref{fig:cuts} that the adopted selection criteria, although highly efficient in eliminating regular disk galaxies and ellipticals, still select mostly galaxies that, although morphologically disturbed, are not necessarily undergoing RPS.  In fact less than 20\% of the automatically selected systems are classified as RPS candidates in our visual screening process.  The disturbed sources rejected after visual inspection can largely be assigned to one of the following classes: strong-gravitational-lensing features (including both cluster-galaxy and galaxy-galaxy lensing events), foreground irregular galaxies, close pairs of ellipticals, unclassifiable clumpy emission in low signal-to-noise areas, and artefacts due to source confusion in crowded regions.  We also note that, while colour information was not directly included in our selection procedure, the availability of images in both the F606W and F814W passbands proved essential in our visual classification to distinguish between the morphological disturbances caused by RPS and irregular extinction due to dust (see Fig.~\ref{fig:need_color}).

\subsection{RPS-candidate sample}
\begin{table*}
  \caption{Properties of the morphologically most compelling "jellyfish" galaxies that constitute our training set. The projected radius $r_{\rm BCG}$ is the projected distance to the (nearest) BCG; the listed angle of incidence is the mean of the values assigned by the three reviewers (see arrows in Fig.~\ref{fig:jellies}. The first six galaxies form the jellyfish sample of ESE.). \label{tab:jellies}}
 \begin{threeparttable}
  \begin{tabular}{cccccccc}
    \hline 
    Name &    $\alpha$ [J2000] & $\delta$ [J2000] & $m_{\rm F606W}$ & $m_{\rm F814W}$ & $r_{\rm BCG}$ [kpc] & Incidence [deg.] & $z$ \\
    \hline 
    \hline 
    MACSJ0257-JFG1 & 02 57 41.4 & $-$22 09 53 & 18.75 & 18.22 & 166 &  10 & 0.3241\\
    MACSJ0451-JFG1 & 04 51 57.3 & $+$00 06 53 & 19.66 & 19.29 & 298 & 50 & 0.4362\\
    MACSJ0712-JFG1 & 07 12 18.9 & $+$59 32 06 & 19.10 & 18.39 & \,\, 87 & 107\,\,\, & 0.3430\\
    MACSJ0947-JFG1 & 09 47 23.1 & $+$76 22 52 & 19.81 & 19.69 & 210 & 34 & 0.3417\\
    MACSJ1258-JFG1 & 12 57 59.6 & $+$47 02 46 & 19.10 & 18.70 & 133 & 45 & 0.3424\\
    MACSJ1752-JFG1 & 17 51 56.1 & $+$44 40 20 & 20.13 & 19.61 & 370 & 120\,\,\, & 0.3739\\
    \hline
    MACSJ0035-JFG1 & 00 35 27.3 & $-$20 16 18 & 19.49 & 19.02 & 182 & 103\,\,\, & 0.3597\\
    MACSJ0257-JFG2 & 02 57 43.5 & $-$22 08 38 & 19.92 & 19.44 & 243 & 130\,\,\, & 0.3297\\
    MACSJ0429-JFG1 & 04 29 33.3 & $-$02 53 02 & 20.97 & 20.64 & 203 & 113\,\,\, & 0.4000\\
    MACSJ0429-JFG1 & 04 29 40.4 & $-$02 53 18 & 20.75 & 20.36 & 334 & 40 & 0.4049\\
    MACSJ0916-JFG1 & 09 16 12.9 & $-$00 25 01 & 20.43 & 19.97 & 334 & 81 & 0.3300\\
    MACSJ1142-JFG1 & 11 42 37.0 & $+$58 31 48 & 20.25 & 19.62 & 549 & 87 & 0.3267\\
    MACSJ1720-JFG1 & 17 20 13.6 & $+$35 37 17 & 20.05 & 19.52 & 309 & 30 & 0.3832\\
    MACSJ1752-JFG1 & 17 52 06.3 & $+$44 40 05 & 20.25 & 20.06 & 747 & 86 & 0.3527\\
    RXJ2248-JFG1   & 22 48 40.2 & $-$44 30 50 & 20.66 & 20.18 & 335 & 64 & 0.3515\\
  \hline
  \end{tabular}  
 \end{threeparttable}
\end{table*}

The process described  in the previous section yielded 223 possible ram-pressure stripping events (including the training set). We consider 15 of these to be classical jellyfish galaxies (yellow symbols in Fig.~\ref{fig:cuts}); an additional 115 objects show characteristic features of RPS (albeit less extreme), and 93 are at least plausible candidates. While we cannot rule out that physical processes other than RPS (e.g., minor mergers or tidal interactions) contribute to, or in fact cause, the observed morphology of our candidates, such alternative scenarios are likely to be relevant mainly for the fainter galaxies in our sample for which the most compelling sign of RPS (evidence of a debris trail) cannot be discerned in the shallow imaging data in hand.

\begin{figure*}
 \centerline{
 \includegraphics[width=0.3\textwidth]{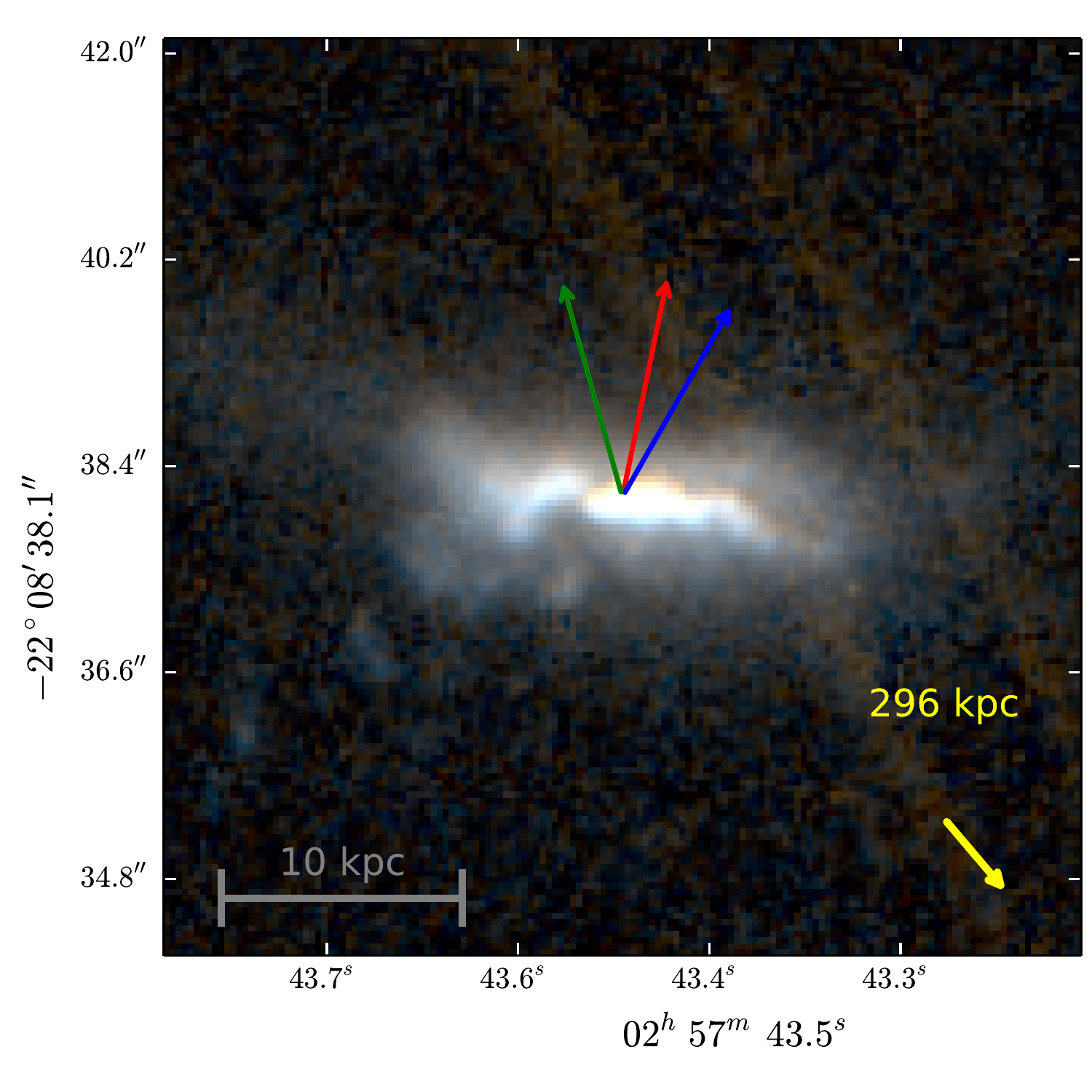}
 \includegraphics[width=0.3\textwidth]{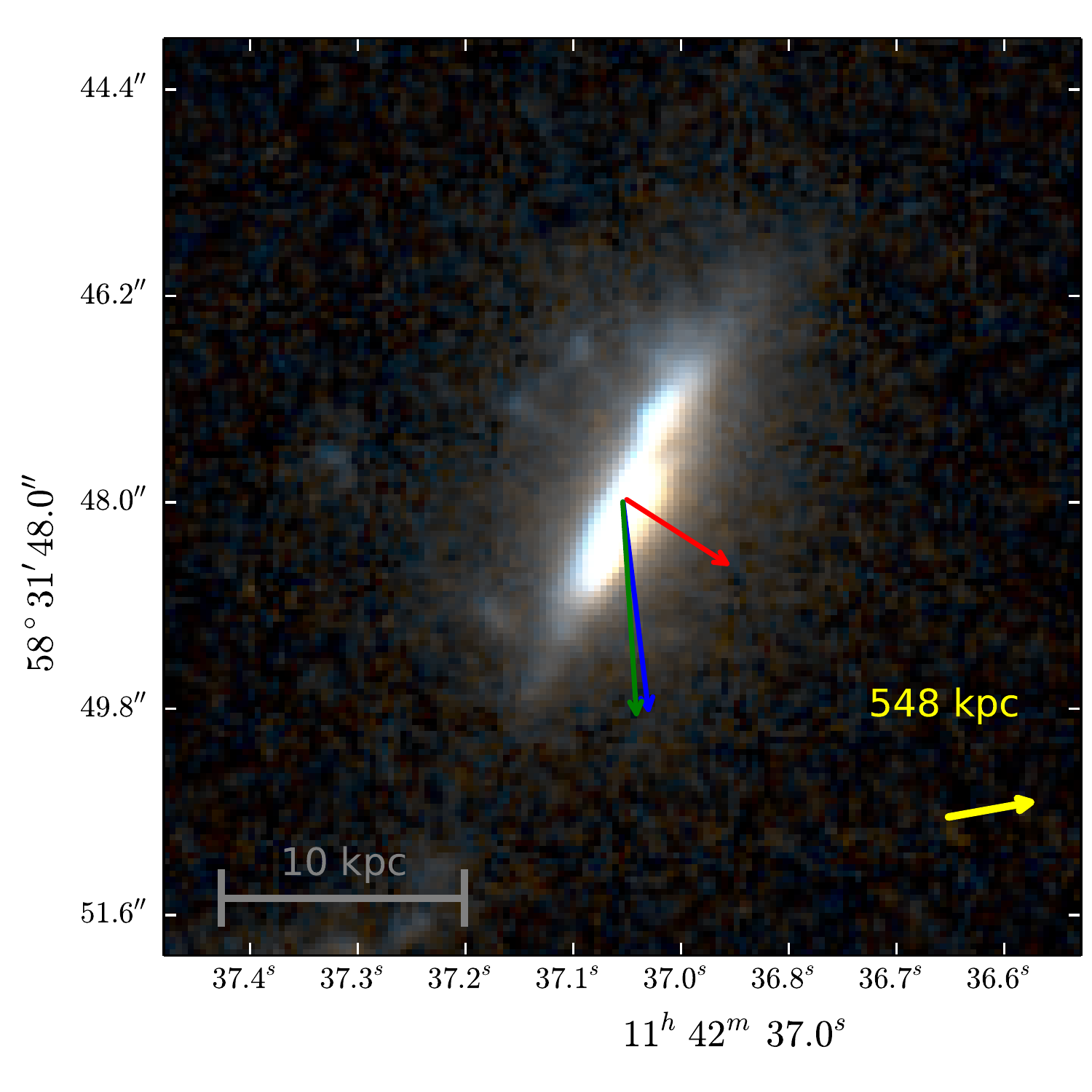}
 \includegraphics[width=0.3\textwidth]{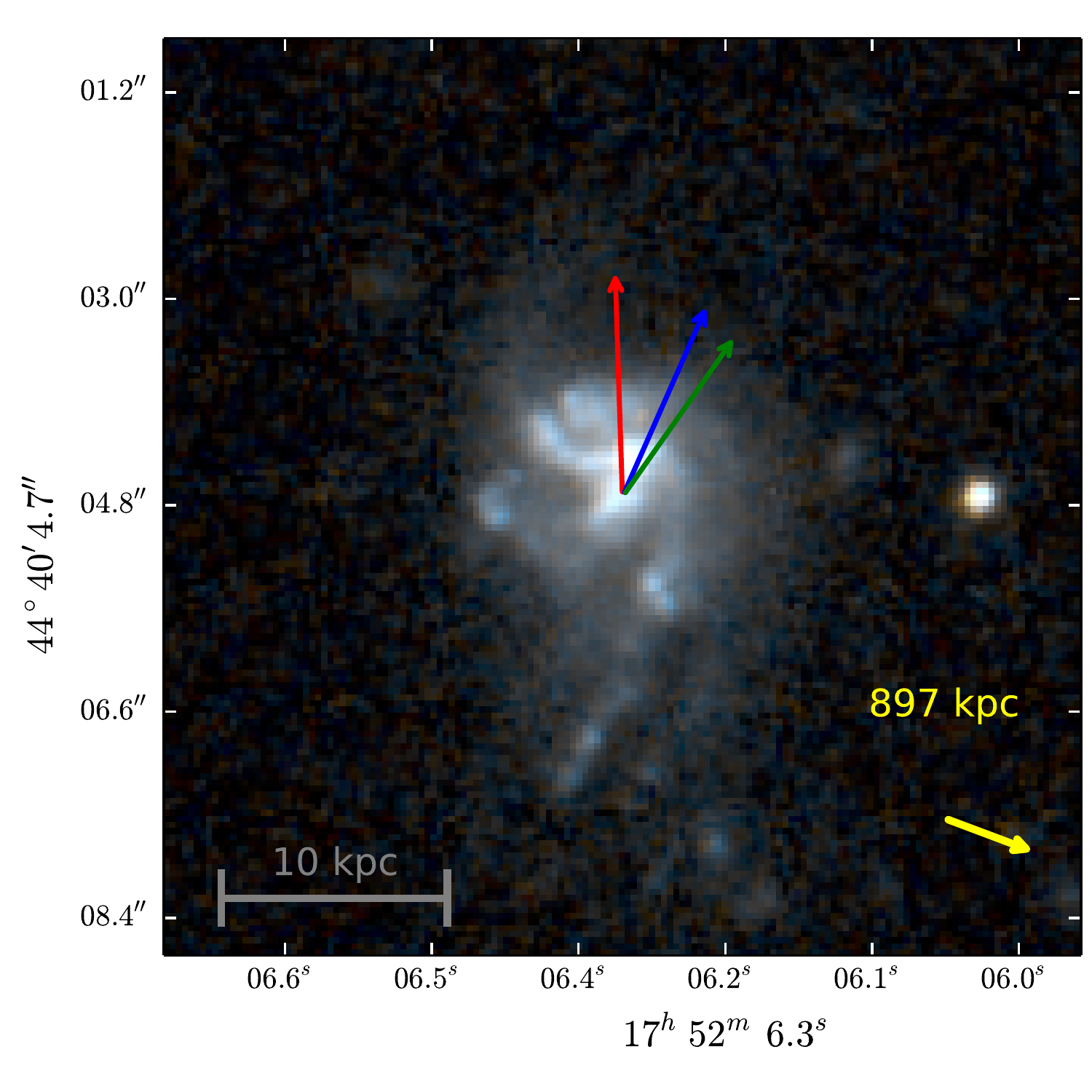}
 }
 \centerline{
 \includegraphics[width=0.3\textwidth]{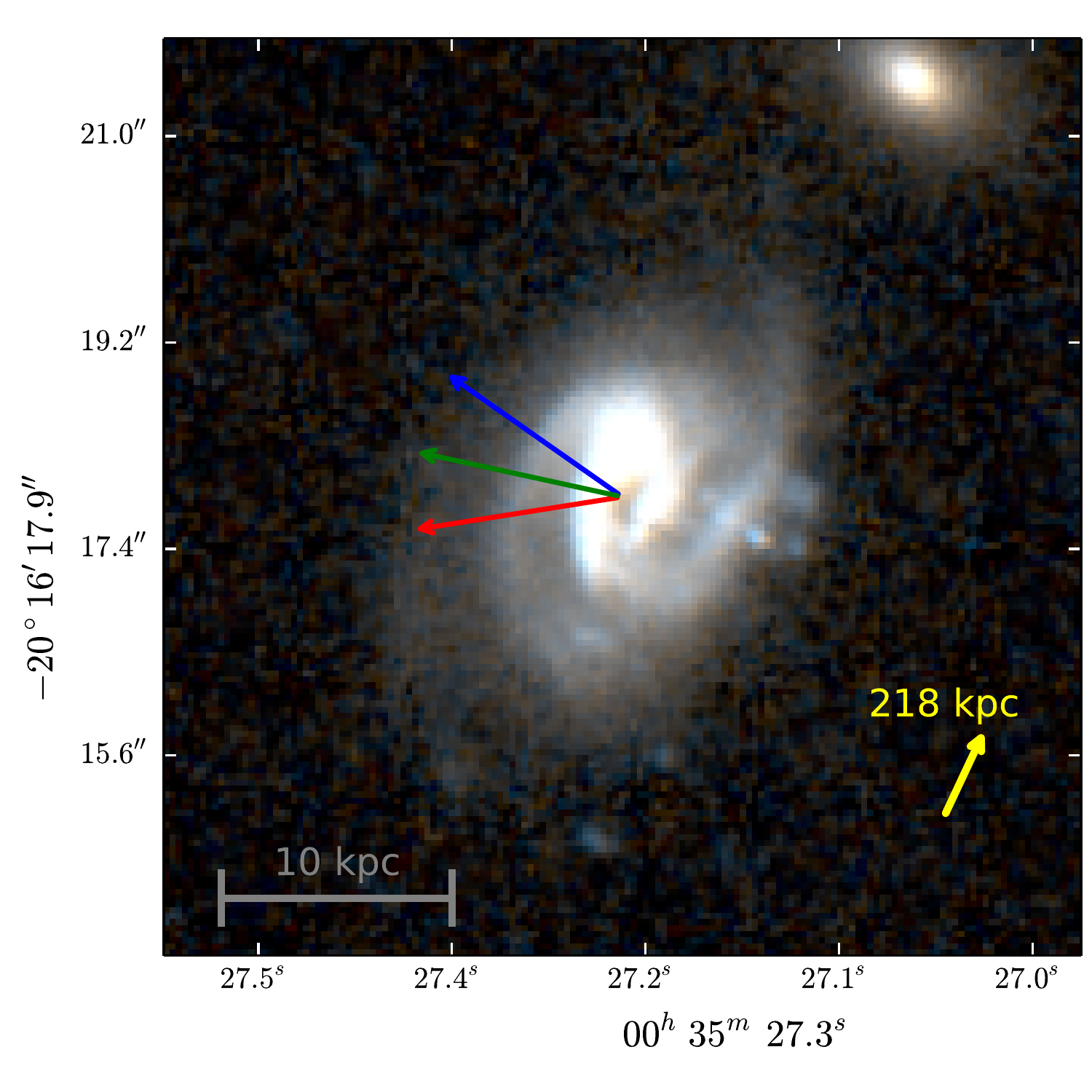}
 \includegraphics[width=0.3\textwidth]{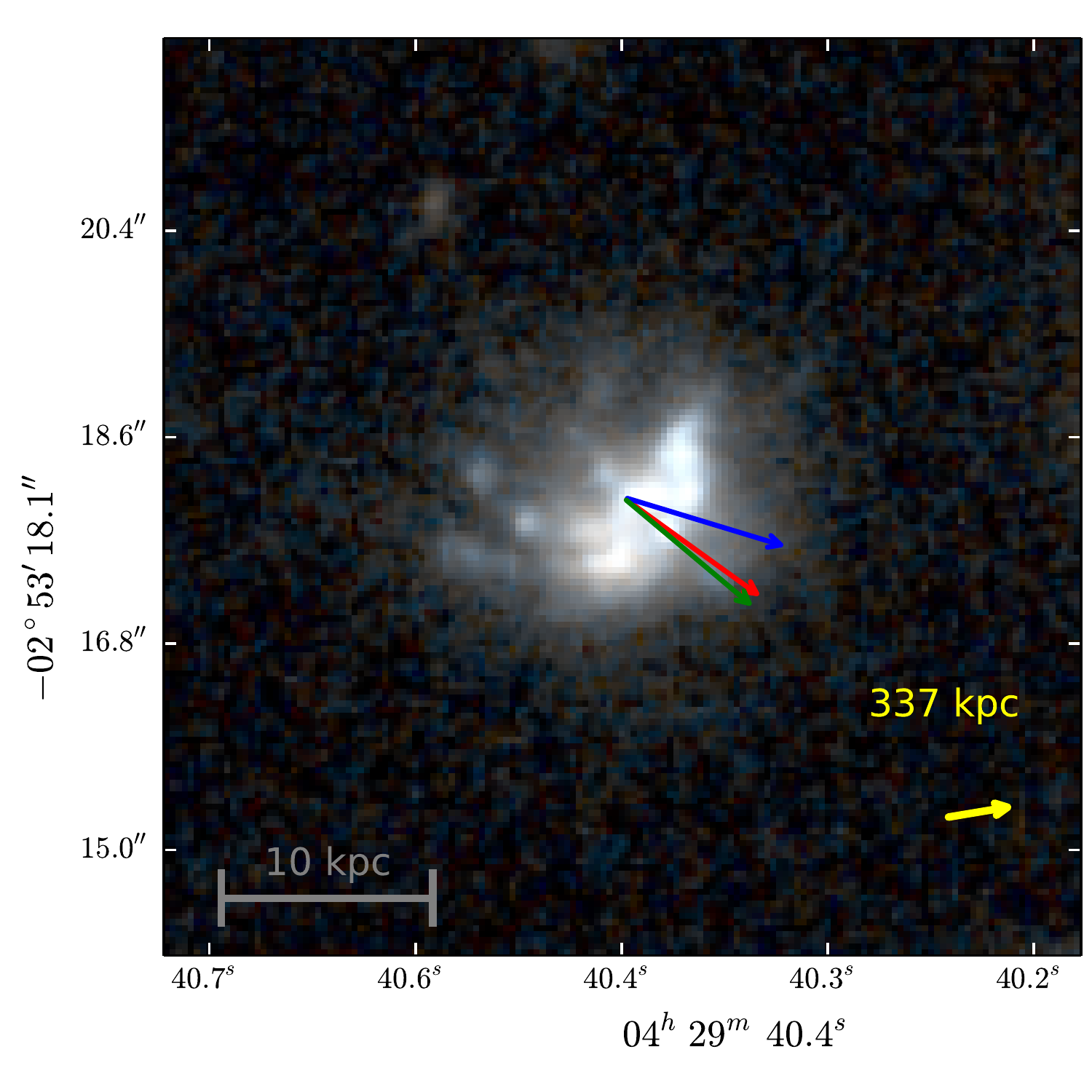}
 \includegraphics[width=0.3\textwidth]{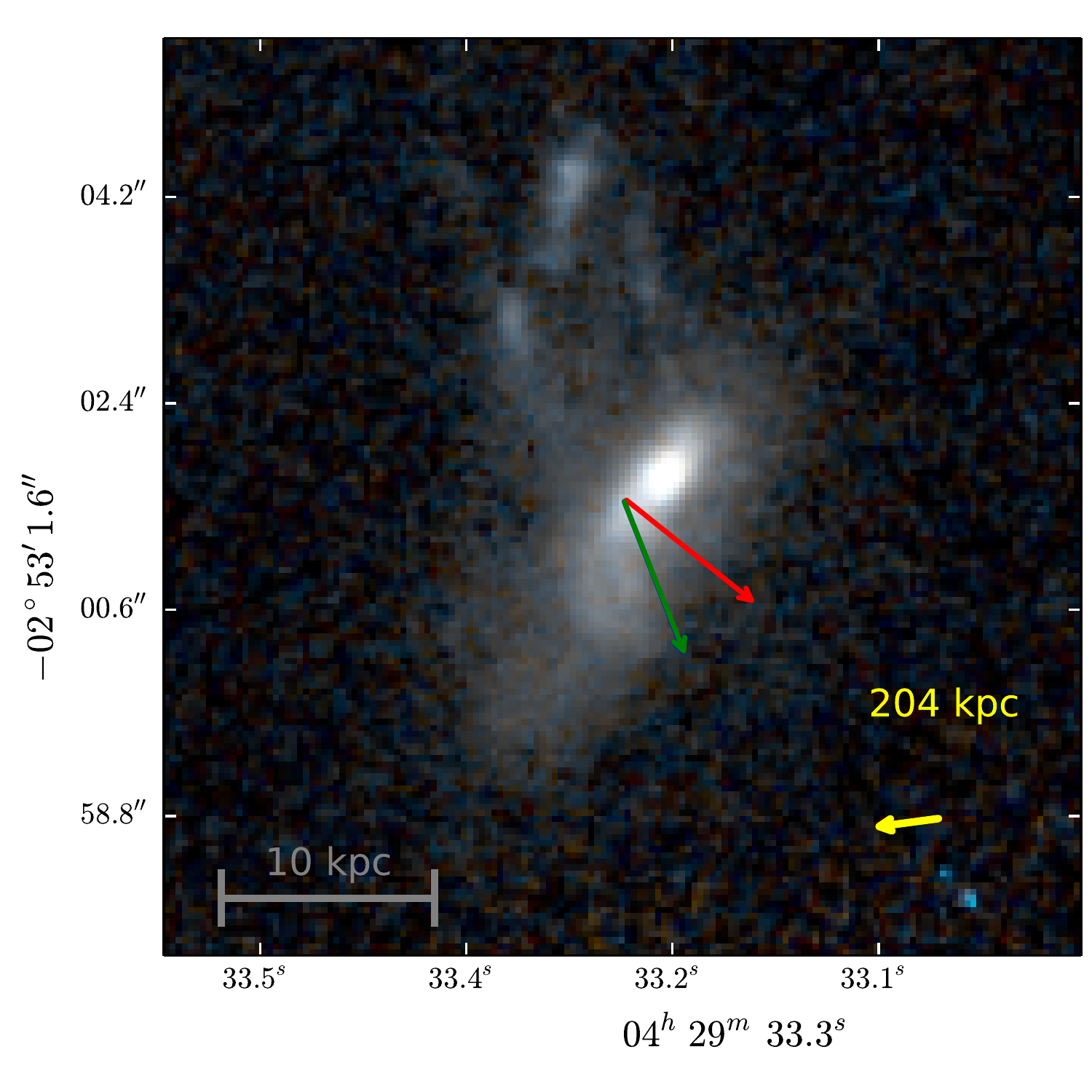}
 }
 \centerline{
 \includegraphics[width=0.3\textwidth]{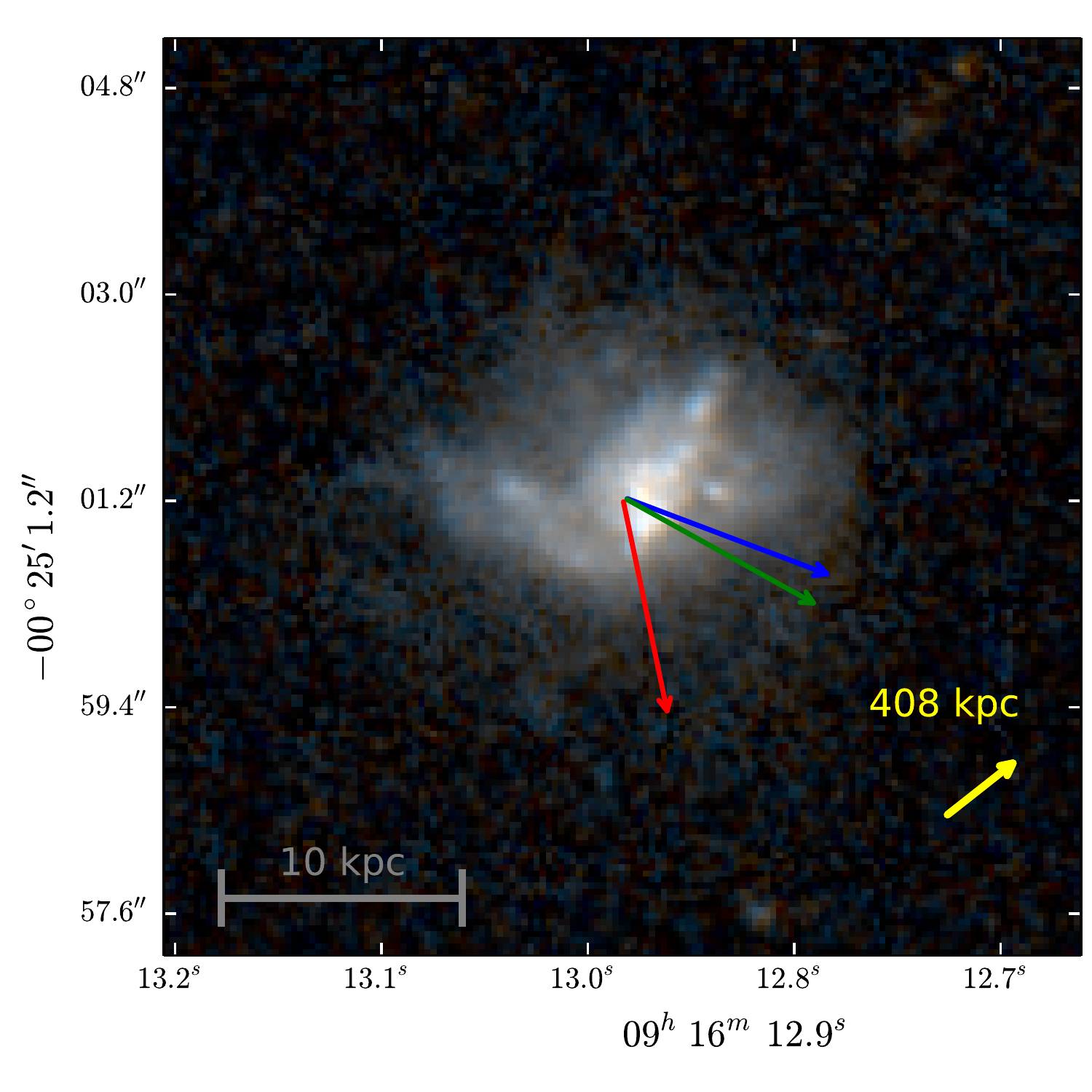}
 \includegraphics[width=0.3\textwidth]{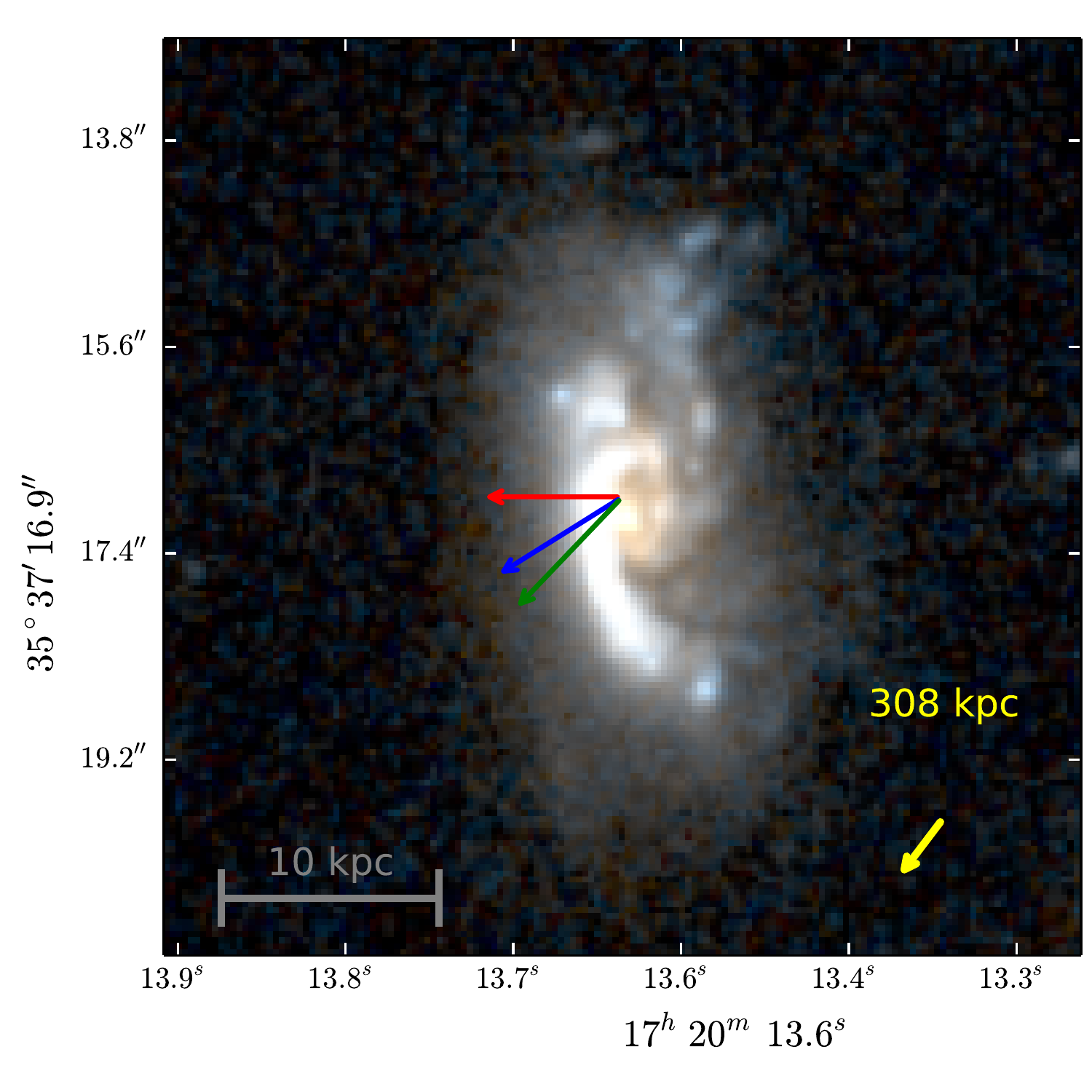}
 \includegraphics[width=0.3\textwidth]{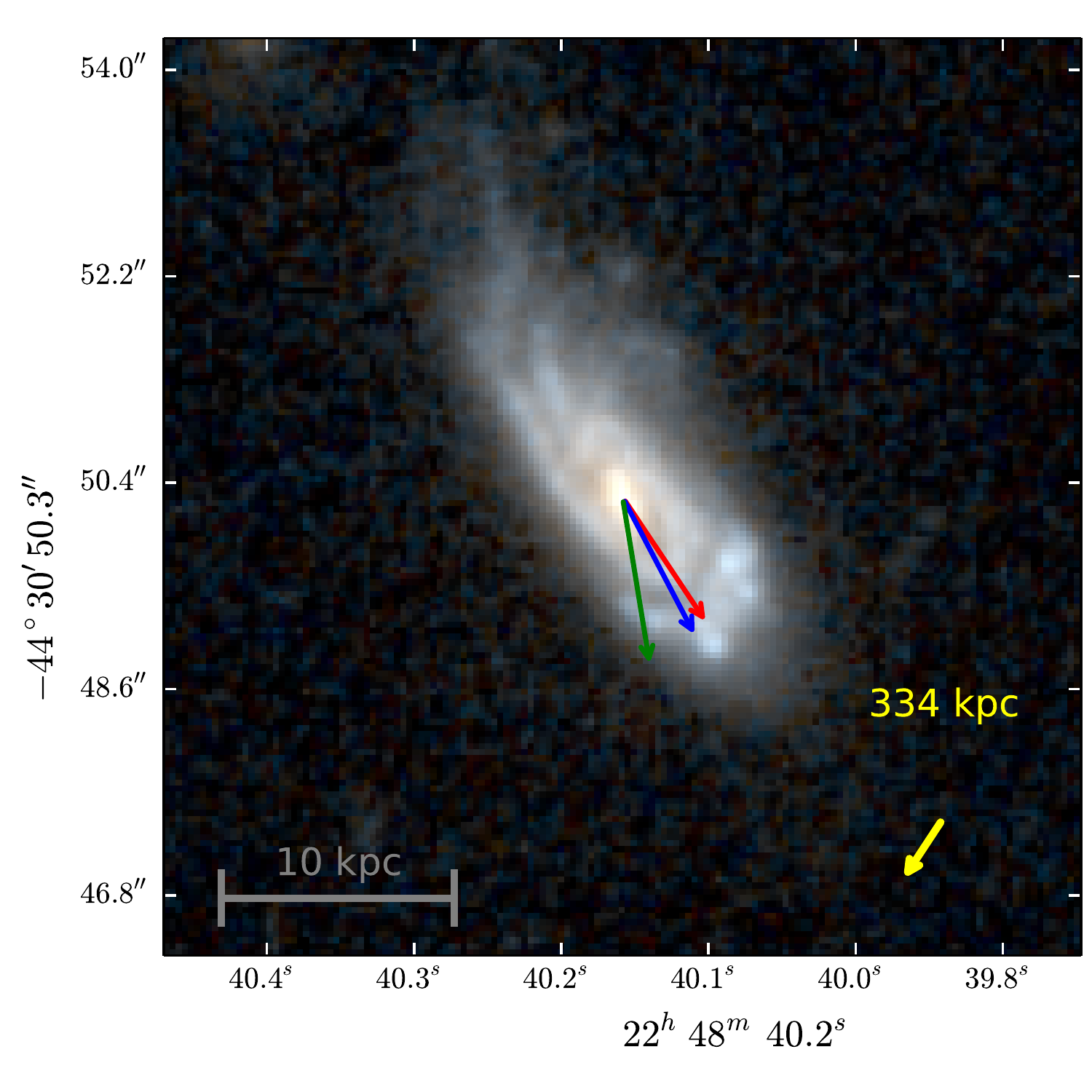}
 }
 \caption{Nine additional textbook examples of ram-pressure stripping discovered in this work; the first three of these were previously identified but not published by \citet{ESE} (see also Fig.~\ref{fig:ese}).   The blue, green, and red arrows indicate the direction of motion assigned to the respective galaxy by the three reviewers; the yellow arrow and metric separation denote the direction and distance to the cluster centre (unknown to the reviewers).\label{fig:jellies}}
\end{figure*}

As a complement to the first six "jellyfish" galaxies discovered in MACS clusters by \citet{ESE}, we show in Fig.~\ref{fig:jellies} a second sample of nine compelling jellyfish galaxies;  fundamental properties of these systems are further described in Section~\ref{sec:arrows} and listed in Table~\ref{tab:jellies}.

\subsubsection{Observational biases}
\label{sec:bias}

Impressive as the list of 223 RPS candidates may appear, we caution again that most of these galaxies may not even be cluster members, and that, for those that are, the cause of the observed morphological features need not be RPS. In addition, our list is almost certainly incomplete. Two primary observational biases are to blame: (a) our inability to reliably discriminate against non-RPS events solely from morphological data (leading to contamination by non-cluster galaxies) and (b) our inability to identify RPS events in galaxies moving close to our line of sight (leading to incompleteness regarding true RPS events in our target clusters).

First results from a comprehensive spectroscopic survey of all candidates (Blumenthal et al., in preparation) indeed indicate that more than half of the objects we selected are in fact fore- or background galaxies.  The hazards of morphological selection alone are underlined not just by this high percentage of projection effects, but also by the elimination of three members of our extended training set (see bottom row Fig.~\ref{fig:ese}):  the edge-on disk with a stellar tail in MACSJ1236.9+6311 is in the foreground of the cluster, while the dramatically distorted face-on spiral galaxy near the core of MACSJ1652.3+5534 was found to be a background object gravitationally lensed by the massive MACS cluster. The bright blue face-on spiral in MACSJ1731.6+2252, finally, turned out to be a member of a foreground group of galaxies. Although the removal of these three objects from our training set has no effect on our selection criteria, as can be seen from Fig.~\ref{fig:cuts} in which these galaxies are marked by red circles, the misidentification of galaxies we considered "textbook" cases of RPS serves as a warning about the robustness of morphological selection and underlines the need for spectroscopic follow-up observations.

The impact of the second observational bias cannot trivially be quantified by means of additional observations. Galaxies moving close to our line of sight lack the tell-tale debris trail and bow-shock morphology readily apparent for RPS proceeding in the plane of the sky (see Fig.~\ref{fig:numsim_models}) and are thus likely to be missed. We attempt to account for the resulting systematic incompleteness when modelling galaxy trajectories in Section~\ref{sec:models}.

\subsection{Direction of motion and location within the cluster}
\label{sec:arrows}

\begin{figure}
 \centerline{
\includegraphics[width=0.47\textwidth]{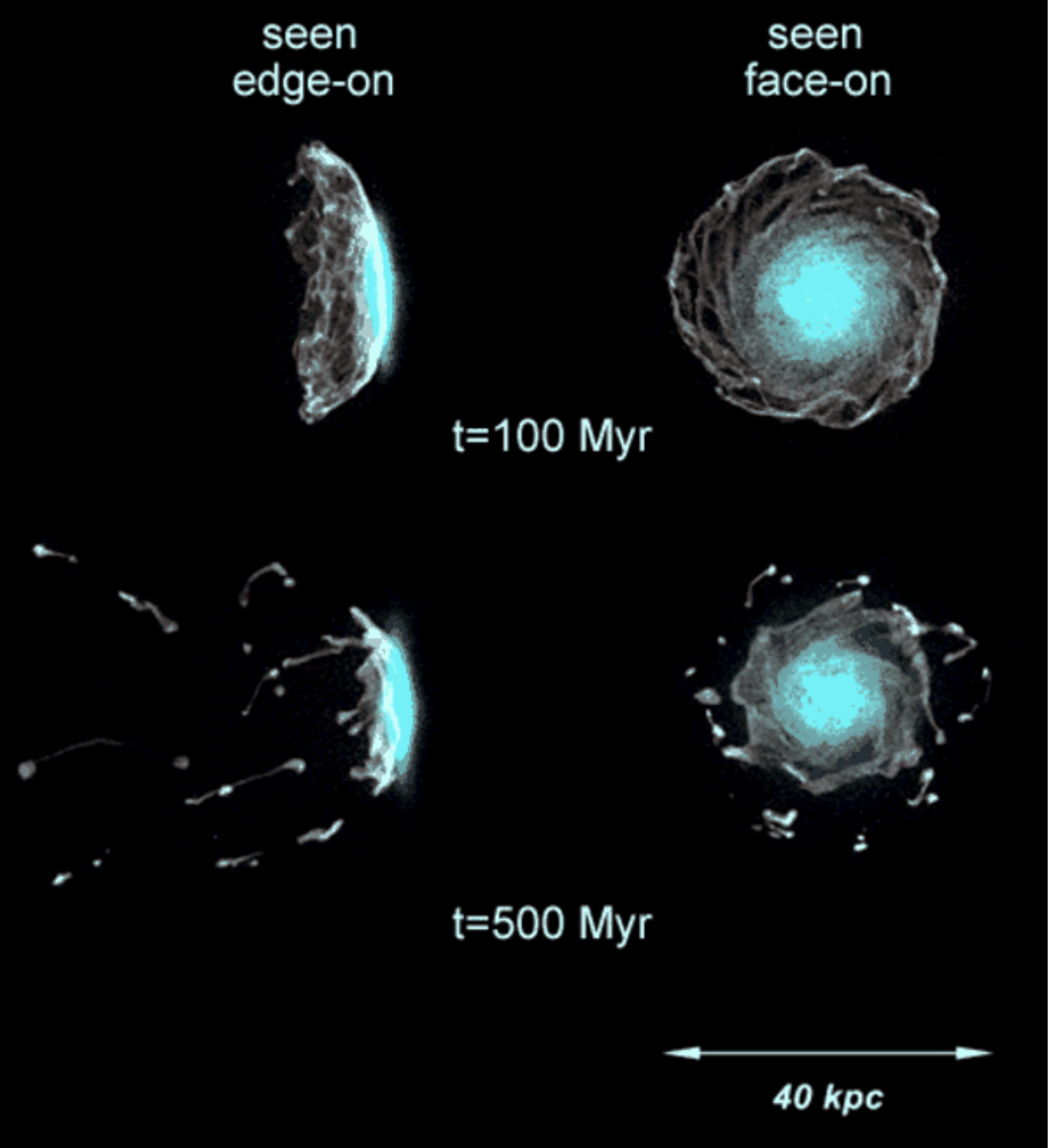}
 }
 \caption{Distribution of gas (white) and newly formed stars (turquoise) for a simulated RPS event involving a spiral galaxy moving face-on through the ICM.  A comparison with Fig.~\ref{fig:ese} shows that our morphological selection is, unsurprisingly,  most sensitive to features typical of mature RPS events in galaxies viewed edge-on.  \citep[Reproduced from][]{kronberger_2008}. \label{fig:numsim_models}}
\end{figure}

Since one of the goals of our study is to distinguish between the different geometric and kinematic scenarios associated with 
"stream-fed" infall along filaments, and cluster mergers,  we focus on two key properties of cluster galaxies: the angle of incidence of their trajectory with respect to the gravitational centre of the cluster and the distance from the cluster centre. To observationally constrain the former, we consult the results of  hydrodynamical modeling of RPS \citep[e.g.,][]{roediger_2006, kronberger_2008, roediger_2014} for insights regarding the correlation between the morphological disturbances caused by RPS and the galaxy's direction of motion. 
Figure \ref{fig:numsim_models} shows model predictions for the distribution of gas and newly formed stars in galaxies undergoing RPS while moving face-on through the ICM. As expected, identifying the direction of motion becomes challenging when a galaxy moves through the ICM along our line of sight or is observed early in the stripping process.

\begin{figure}
 \centerline{
 \includegraphics[width=0.23\textwidth]{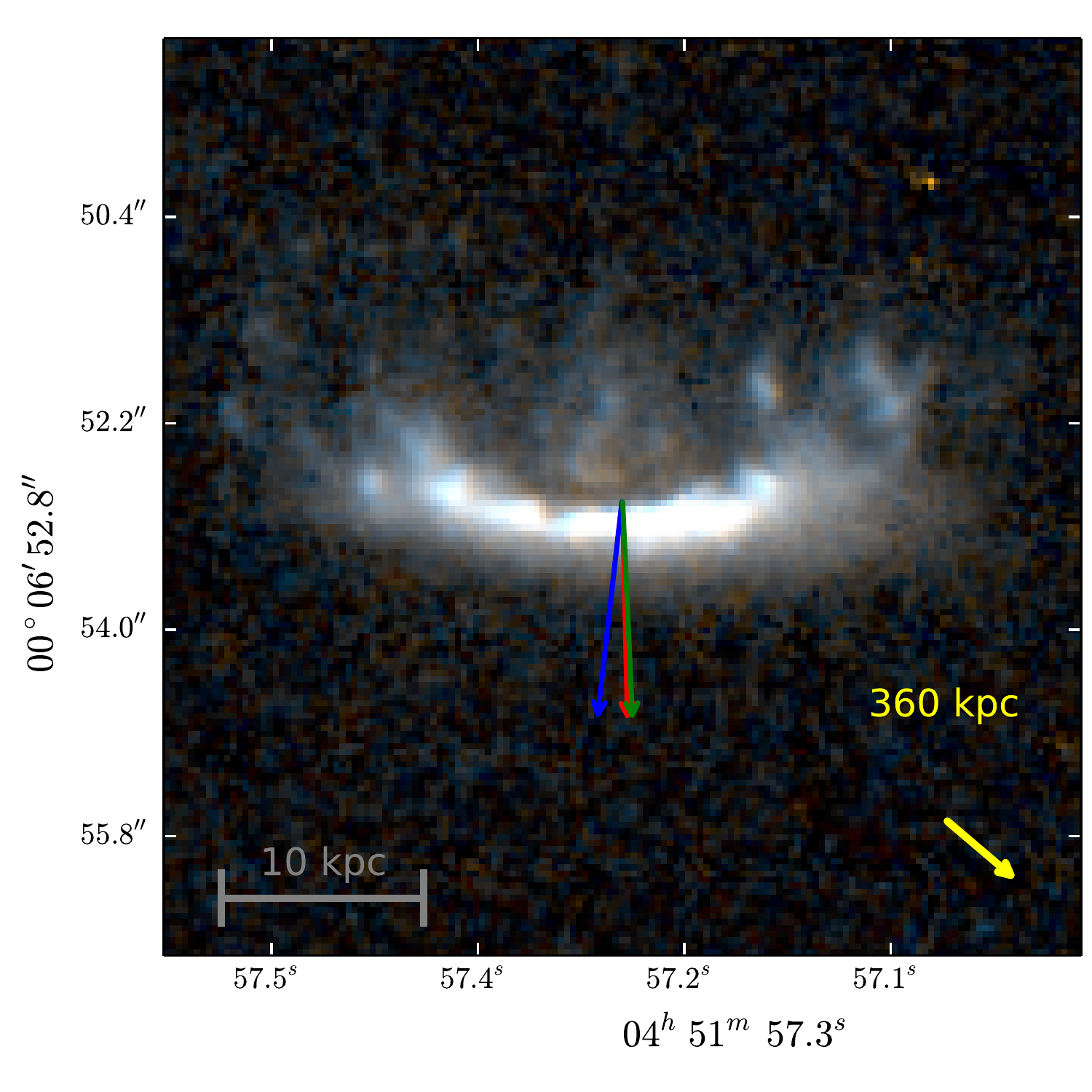}
 \includegraphics[width=0.23\textwidth]{macsj0429_all_id_298.pdf}
 }
 \centerline{
 \includegraphics[width=0.23\textwidth]{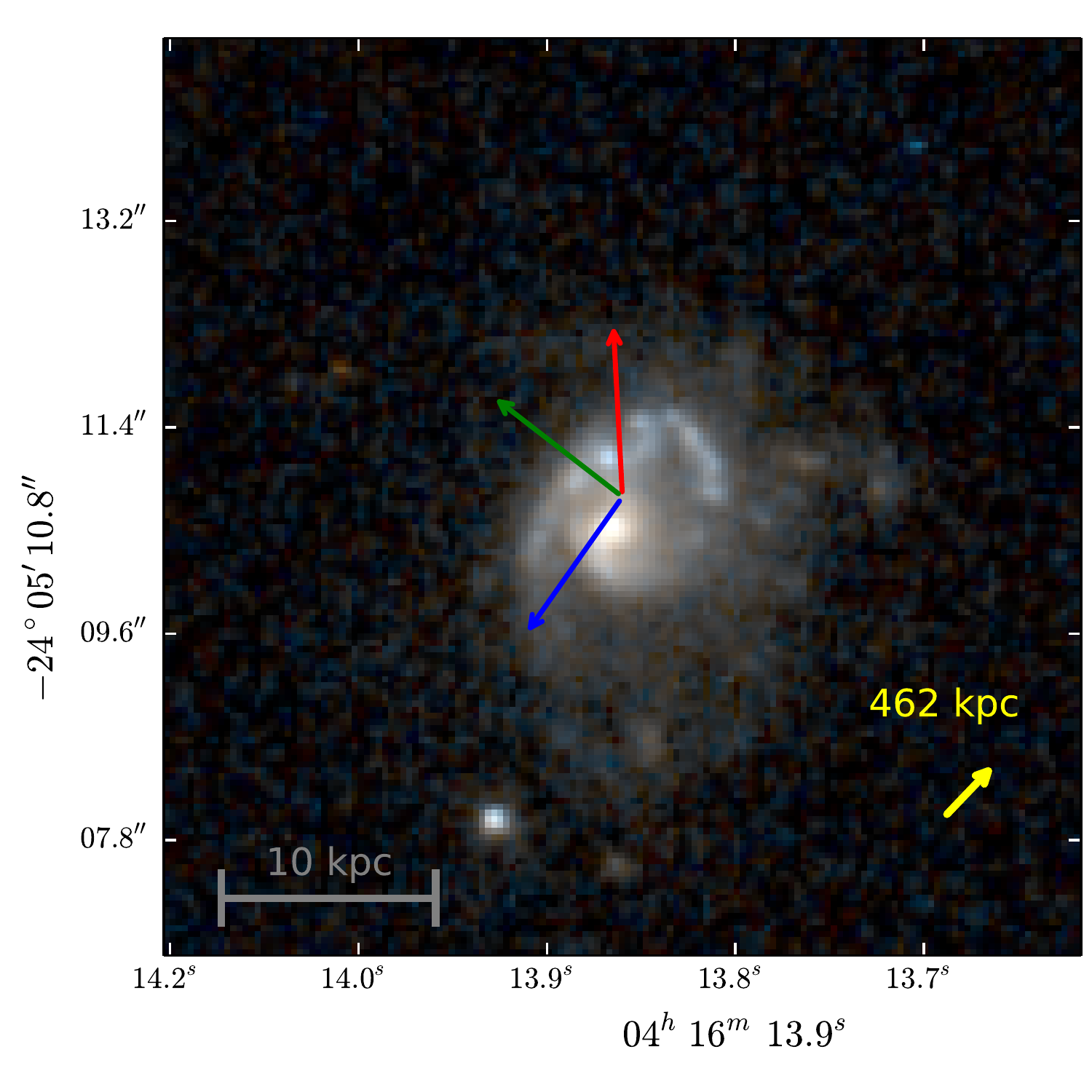}
 \includegraphics[width=0.23\textwidth]{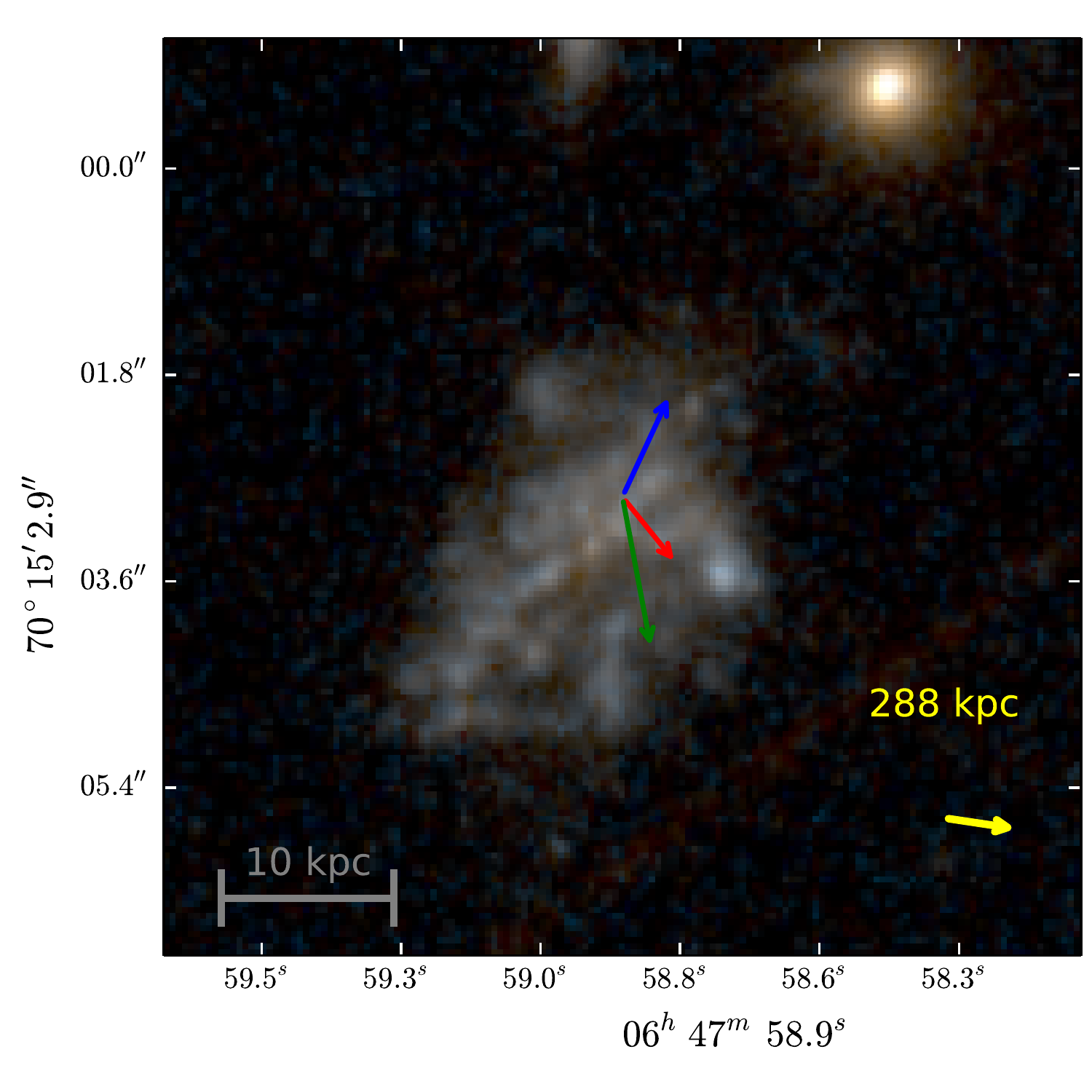}
 }
 \caption{Examples of RPS candidate events illustrating our process to estimate direction of motion and the associated error. The arrows are the same as in Fig.~\ref{fig:jellies}. \label{fig:arrows}}
\end{figure}

We attempt to assign projected directions of motion visually according to the following prescriptions: (1) if tails are discernible, the velocity vector is assumed to be parallel to the tail; (2) edge-on disks showing significant curvature are assigned velocity vectors oriented perpendicular to said curvature and originating at its apex; (3) if extended regions of star formation appear to be present, the velocity vector is placed perpendicular to the dominant elongation of said regions; (4) if none of the previously mentioned indicators are present (or if they are contradictory), we attempt to make the best physically motivated estimate.  To avoid systematic biases, galaxies are inspected using small thumbnail images covering only the region immediately surrounding the galaxy with no indication of the direction to the cluster centre.  In recognition of the subjective nature of our visual measurements \citep[especially for galaxies moving partly or largely along our line of sight,][]{roediger_2006}, the process is performed independently by three reviewers to derive an approximate grade for the robustness of each estimated direction of motion. Figure~\ref{fig:arrows} shows examples of objects falling into each of our quality grades with uncertainty increasing top to bottom and left to right.  We then define the angle of incidence as the angle between the apparent velocity vector and the position vector with respect to the cluster centre (taken to be the location of the brightest cluster galaxy, BCG), i.e., the angular deviation from a purely radial infall trajectory (note again that all of these quantities are defined and observed in projection).  

A second galaxy property that is critical to our efforts to deduce trajectories is location within the cluster.  For RPS candidates  lacking radial-velocity measurements, we are unable to assess whether an object is located in front or behind the cluster centre (defined by the redshift of the BCG), let alone further constrain its physical distance to the latter along our line of sight.  Projected distances, however, measured in the plane of the sky and relative to the location of the BCG, are trivially obtained for comparison with the distribution expected for different geometries of galaxy infall. 
 
\section{A simple model of galaxy trajectories}
\label{sec:models}

In order to understand which kind of galaxy trajectories are most compatible with the observed distributions of (projected) incidence angle and cluster-centric distance, we compare our observations with the results of a simple theoretical model. To this end, we calculate orbits in a canonical cluster representative of the MACS clusters in our sample and use simple prescriptions, described below, to predict the projected radii and incidence angles at which extreme RPS events are most likely to occur.

As an infalling galaxy approaches the dense cluster core, the ICM exerts an increasing ram-pressure, $p_\mathrm{ram} = \rho_\mathrm{ICM}v_\mathrm{gal}^2$, where $\rho_\mathrm{ICM}$ is the ICM mass density and $v_\mathrm{gal}$ is the relative velocity between the galaxy and ICM \citep[][hereafter GG]{gunn_infall_1972}. By comparing $p_\mathrm{ram}$ to the gravitational restoring force per unit area on the gas within the galaxy,
\begin{equation}
f_\mathrm{grav}(R) = \Sigma_\mathrm{gas}(R)\frac{\partial\Phi}{\partial Z}(R),
\end{equation}
we find the critical radius where $p_\mathrm{ram} = f_\mathrm{grav}(R_\mathrm{strip})$ \citep{roediger_2007}. Here $\Sigma_\mathrm{gas}$, $\Phi$, and $Z$ are the ISM mass surface density, the gravitational potential of the galaxy, and its scale height, respectively. Beyond $R_\mathrm{strip}$, the galaxy potential is not strong enough to retain the gas and stripping sets in. \citet{vollmer_2001} give an analytic estimate for the GG criterion which determines the stripping radius:
\begin{equation}
\Sigma_\mathrm{gas} v_\mathrm{rot}^2 R_\mathrm{strip}^{-1} = p_\mathrm{ram},
\end{equation}
where $v_\mathrm{rot}$ is the rotation speed of the galaxy. Although, in reality, the onset of RPS is likely to be a highly non-linear process, the simple GG criterion has proven suitable for global characterisations of RPS in in-depth numerical simulations \citep[e.g., ][]{roediger_2007, kronberger_2008}.

\subsection{Galaxy properties}
Since our simple model aims only to predict the distribution of RPS events along galaxy orbits, but not the detailed properties of such events, we model all galaxies in our simulation as thin disks with radius $R_\mathrm{gal} = 15$ kpc and gas surface density $\Sigma_\mathrm{gas}$ = $10^{21}$ atoms per cm$^{2}$ moving face-on through the ICM. 

To account for galaxy-to-galaxy variation in $f_\mathrm{grav}$, we also run all models for a range of galaxy masses, parametrized by the rotational velocity $v_\mathrm{rot}$ (see Eq.~2). The explored range of $v_\mathrm{rot}$ from 150 to 350 km s$^{-1}$ corresponds to dynamical masses, within 15 kpc, of $8\times10^{10}$, $2\times10^{11}$, and $4\times10^{11}$ M$_\odot$. The adopted range of rotational velocities covers a spectrum of masses from sub- to super-Milky Way sized objects. 

\subsection{Cluster properties}
We describe the gas and total mass distribution within the cluster using a spherical $\beta$-model \citep{cavaliere_1976}
\begin{equation}
 \rho = \rho_0\left[1 + \left(\frac{r}{r_0}\right)^2\right]^{-\frac{3}{2}\beta },
\end{equation}
where $\rho_0$ is the central mass (or gas) density, $\beta$ and $r_0$ are the power-law index and core radius, respectively, and $r$ is the cluster-centric radius.
We adopt a total mass of $1.3\times10^{15} M_\odot$ \citep[the average weak-lensing mass, $M(r<1.5$ Mpc$)$, of MACS clusters at $z>0.3$ according to][]{applegate_2014}. As the majority ($\sim$2/3) of the clusters in our sample do not show dramatic large-scale substructure, we assume that our model cluster is largely relaxed, featuring gas and total mass distributions that share a common centre, core radius $r_0$ and power-law slope $\beta$. We adopt $r_0$ = 180 kpc and $\beta$ = 0.59, the median of the values from the spatial X-ray analysis of \citet{mantz_2010}. Assuming a gas fraction $f_{\rm gas}=0.074$ \citep{mantz_2014} and the model parameters above, our canonical cluster has a central particle density $n_0$ of $2.29\times10^{-3}$ cm\textsuperscript{-3}.

\subsection{Galaxy trajectories}

\begin{figure}
 \centerline{
  \includegraphics[width=0.5\textwidth]{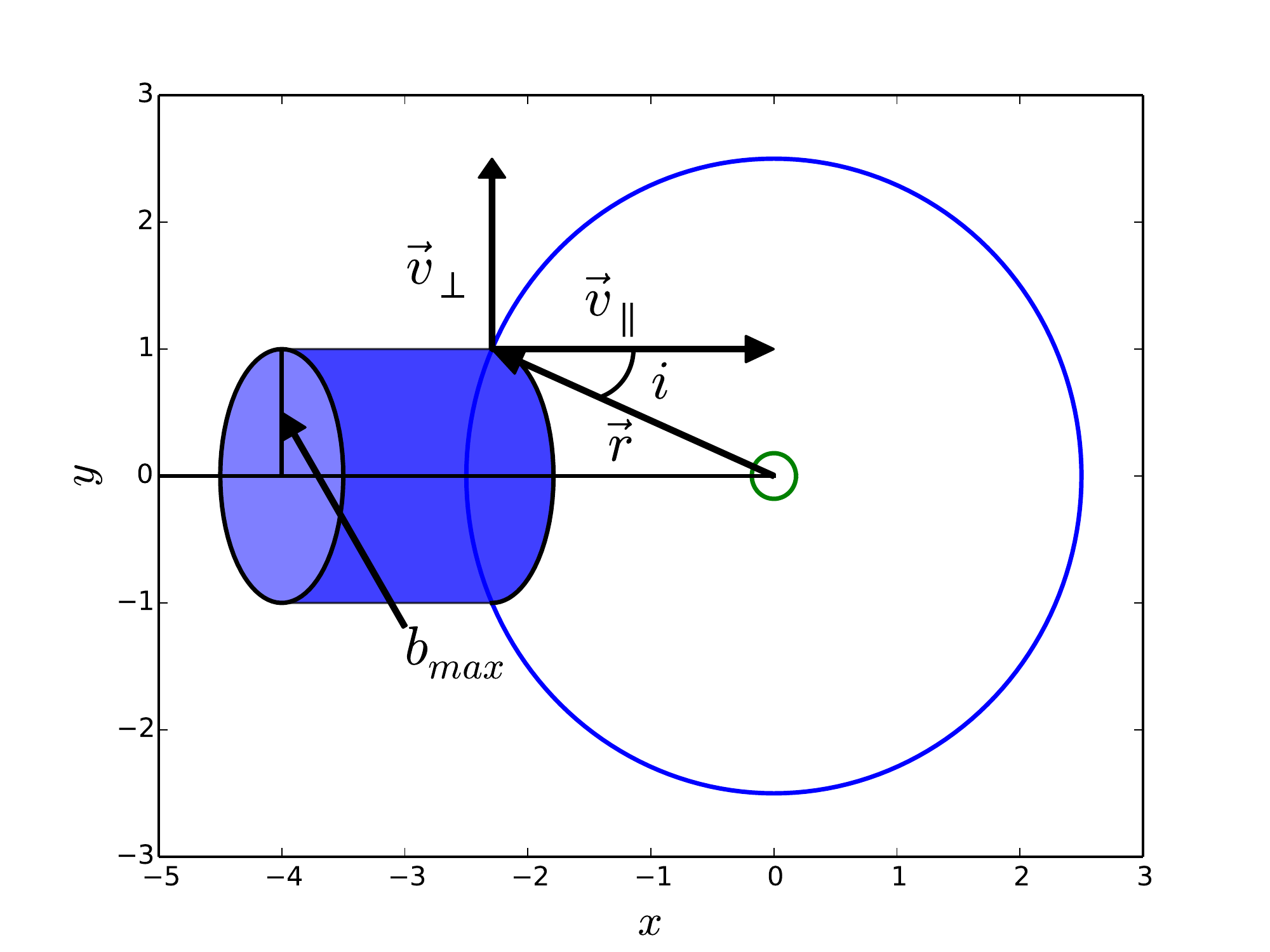}
 }
 \caption{Schematic diagram of the quantities that characterise the initial conditions and orbits of galaxies in our infall models: the maximal impact parameter  $b_\mathrm{max}$, the initial velocities $v_\parallel$ and $v_\perp$, the cluster-centric radius $\hat{r}$, and the inclination angle $i$. \label{fig:vectors}}
\end{figure}

The orbits of test particles falling into our model cluster are computed for a wide range of initial orbital parameters that encompass expectations for infall along connected filaments and from cluster mergers. Orbit calculations begin at the end of a filament which is assumed to be at a distance of 2.5 Mpc from the cluster core ($\approx R_\mathrm{vir}$). In Fig.~\ref{fig:vectors}, we show a schematic of the quantities that characterise orbits in our model: the speed of a galaxy in the direction of the filament axis $v_\parallel$, the transverse velocity perpendicular to the filament flow $v_\perp$, and the impact parameter $b$. 

Radial profiles of filaments in cosmological simulations show a well defined edge at a radius of 1.0--2.0 $h^{-1}$ Mpc ($\sim$1.4--2.8 Mpc in our assumed cosmology) beyond which the matter density essentially vanishes \citep{colberg_2005}. We therefore model filaments as cylinders of constant density with radius $b_\mathrm{max}$. We populate these filaments with $3\times10^4$ galaxies using Monte Carlo sampling designed to provide constant density within $b_\mathrm{max}$ and a normal distribution in $v_\perp$ to account for the velocity dispersion of galaxies within the filament.

In the following, we consider three infall scenarios that differ primarily in the approach velocity of galaxies at the cluster's virial radius: 1) \emph{stream-fed} infall along filaments; 2) a \emph{slow merger}; and 3) a \emph{fast merger}. Table~\ref{tab:models} lists the model parameters that characterise each of these scenarios. For each infall scenario, we fix the initial velocity $v_\parallel$ at one value for all orbits. For the stream-fed model, we choose $b_{\rm max}$=1.5 Mpc and $v_\parallel$ = 200 km s$^{-1}$, the average filament radius and the average velocity of matter at the cluster-filament interface, respectively \citep{colberg_2005}, as well a velocity dispersion characteristic of group environments ($\sim$100 km s$^{-1}$). The slow and fast merger models are characterised by initial velocities of 1000 and 3000 km s$^{-1}$, respectively, and a velocity dispersion of 1000 km s$^{-1}$ and $b_\mathrm{max}$ = 2.5 Mpc for either merger scenario. Since we know neither the number and orientation of connected filaments for our cluster sample, nor the orientation of a putative merger axis, we place filaments/merging clusters at $10^3$ positions, sampled isotropically on a 2.5 Mpc sphere. In total, this results in $3\times10^6$ orbits per scenario which are each followed for 5 Gyr ($\sim t_\mathrm{cross}$) in time steps of 5 Myr. 

\begin{table}
  \caption{Model Parameters\label{tab:models}}
 \begin{threeparttable}
  \begin{tabular}{lccc}
    \hline 
    Model & $v_\parallel$ [km s$^{-1}$] & $\sigma_v$ [km s$^{-1}$] & $b_\mathrm{max}$ [Mpc] \\
    \hline 
    \hline 
    Stream-fed  & 200  & 100  & 1.5 \\
    Slow Merger & 1000 & 1000 & 2.5 \\
    Fast Merger & 3000 & 1000 & 2.5 \\
  \hline
  \end{tabular}  
  \begin{tablenotes}
   \item See Fig.~\ref{fig:vectors} for a schematic illustration of $v_\parallel$ and $b_\mathrm{max}$; $\sigma_v$ indicates the velocity dispersion of infalling galaxies.
  \end{tablenotes}
 \end{threeparttable}
\end{table}

Defining the start of the RPS event as the time step in which the GG criterion is first satisfied, we explore a range of RPS event durations, from 50 Myr to 1 Gyr, during which the resulting event is assumed to remain observationally detectable.  This choice is motivated by numerical simulations: \citet{roediger_2014} find that the signature RPS morphology should be observable in galaxies overrun by an ICM shock for between $\sim$several 10 Myr to a few 100 Myr. Slightly longer durations are quoted by \citet{kronberger_2008} for a scenario similar to our stream-fed infall model (see also Fig.~\ref{fig:numsim_models}).

For comparison with our observational results, segments of the orbits corresponding to an RPS event (under our definition) are projected onto the plane of the sky, thus providing the projected angle of incidence (the projected angle between the galaxy's velocity and position vectors), $i$, and the projected radius from the cluster centre. We then tabulate the amount of time spent in bins of projected radius and inclination angle to construct simulated probability distributions for each scenario.

\subsection{Accounting for observational bias}
\label{sec:bias2}

As mentioned in Section~\ref{sec:bias} and illustrated in Fig.~\ref{fig:numsim_models}, RPS events in galaxies moving along or close to our line of sight are likely to be missed, as, for this particular geometry, the pronounced morphological features that our selection process is build upon are obscured by the galaxy being stripped. We examine the importance of this observational bias by imposing on our modeling results that all RPS events are undetectable that occur in galaxies moving along an axis that is inclined to our line of sight by 0, 15, 30, or 45 degrees. As detailed in the following section, even the most severe implementation of this line-of-sight bias results in only modest changes in the model predictions, suggesting that the effect does not significantly affect the conclusions drawn from our comparison with the data.

\begin{figure*}
  \centerline{
  \includegraphics[width=\textwidth]{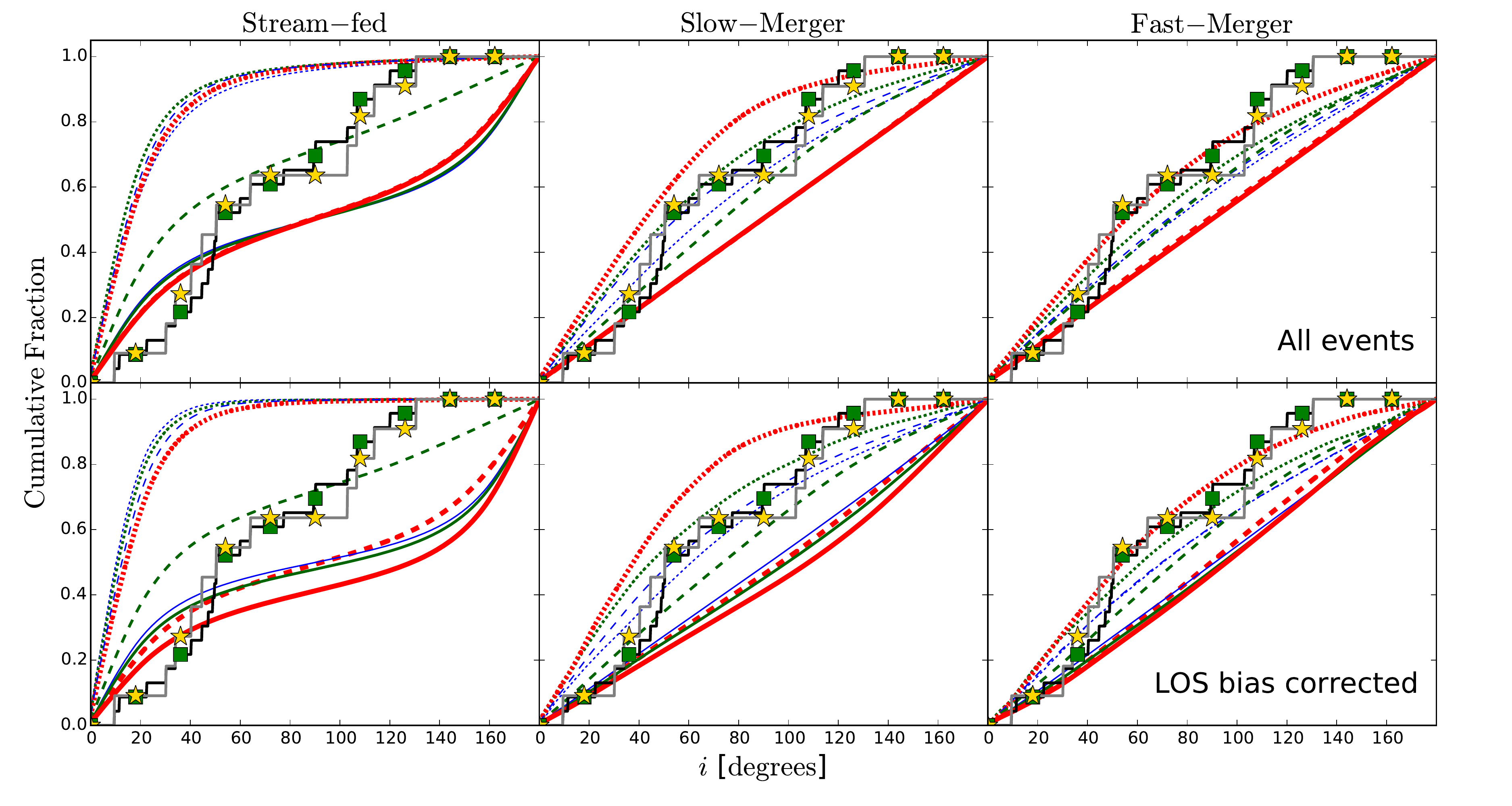}
 } 
 \caption{Cumulative distribution of the incidence angles of the jellyfish and cluster-member samples (asterisks and squares, respectively). The left, center, and right panels show predictions for the stream-fed, slow-merger, and fast-merger models (see Table~\ref{tab:models}), respectively. The dotted, dashed, and solid lines correspond to event durations of 50 Myr, 300 Myr, and 1 Gyr, respectively.  Colors denote the mass of the infalling galaxy:  blue (thin), green (medium), and red (thick) correspond to dynamical masses of $8\times10^{10}$, $2\times10^{11}$, and $4\times10^{11} M_\odot$, respectively. Model predictions shown in the top row assume that RPS events are identifiable as such regardless of the inclination of the galaxy's direction of motion with respect to our line of sight; results shown in the bottom row mimic the observational bias discussed in Sections~\ref{sec:bias} and \ref{sec:bias2} by excluding all events triggered in galaxies with velocity vectors within 30 degrees of our line of sight. \label{fig:incid_cum_dist}}
\end{figure*}

\begin{figure*}
 \centerline{
  \includegraphics[width=\textwidth]{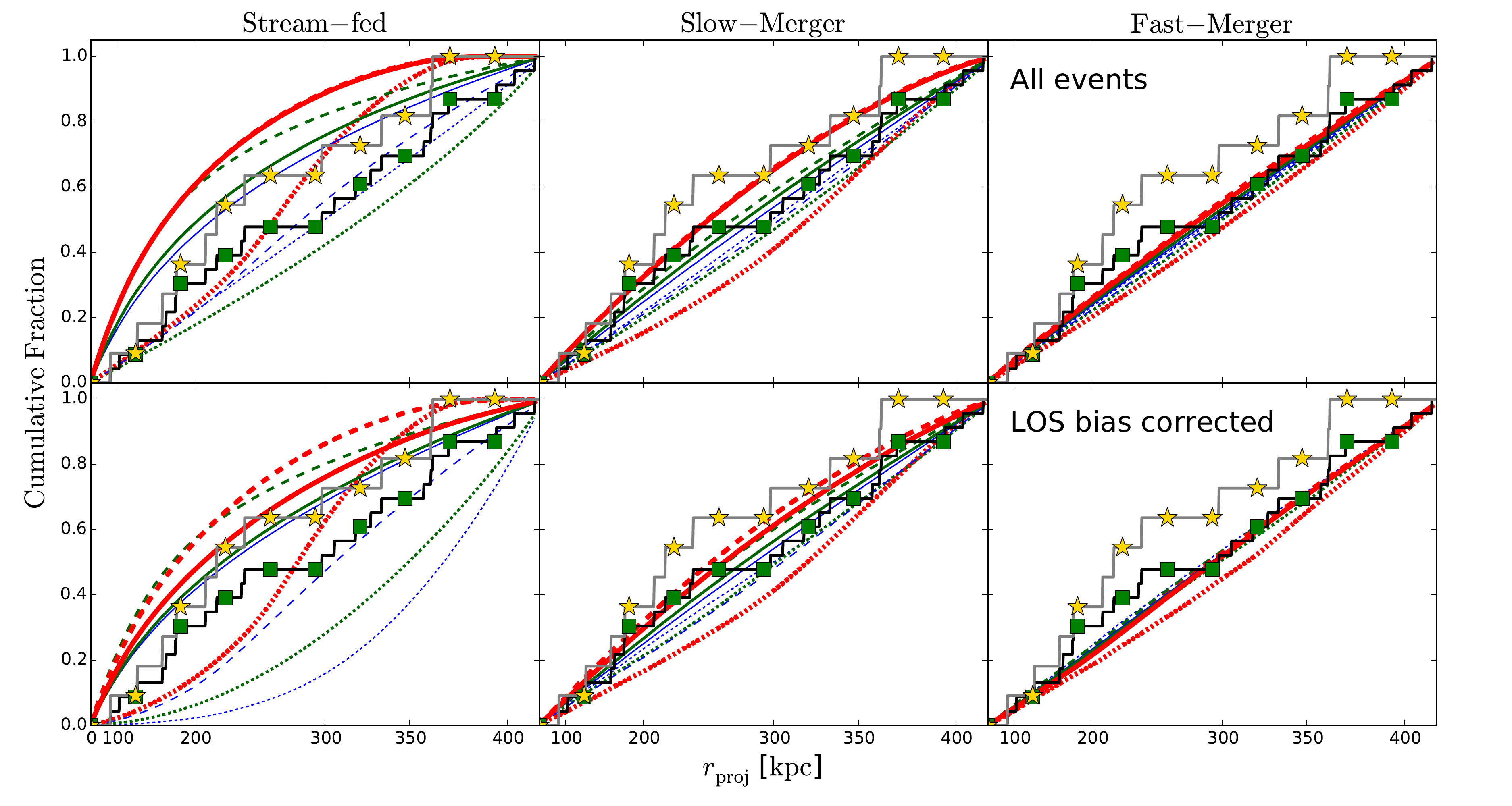}
  }
 \caption{As Fig.~\ref{fig:incid_cum_dist} but for the projected radius, $r_\mathrm{proj}$. \label{fig:rad_cum_dist}}
\end{figure*}

\begin{figure*}
 \centerline{
  \includegraphics[width=\textwidth]{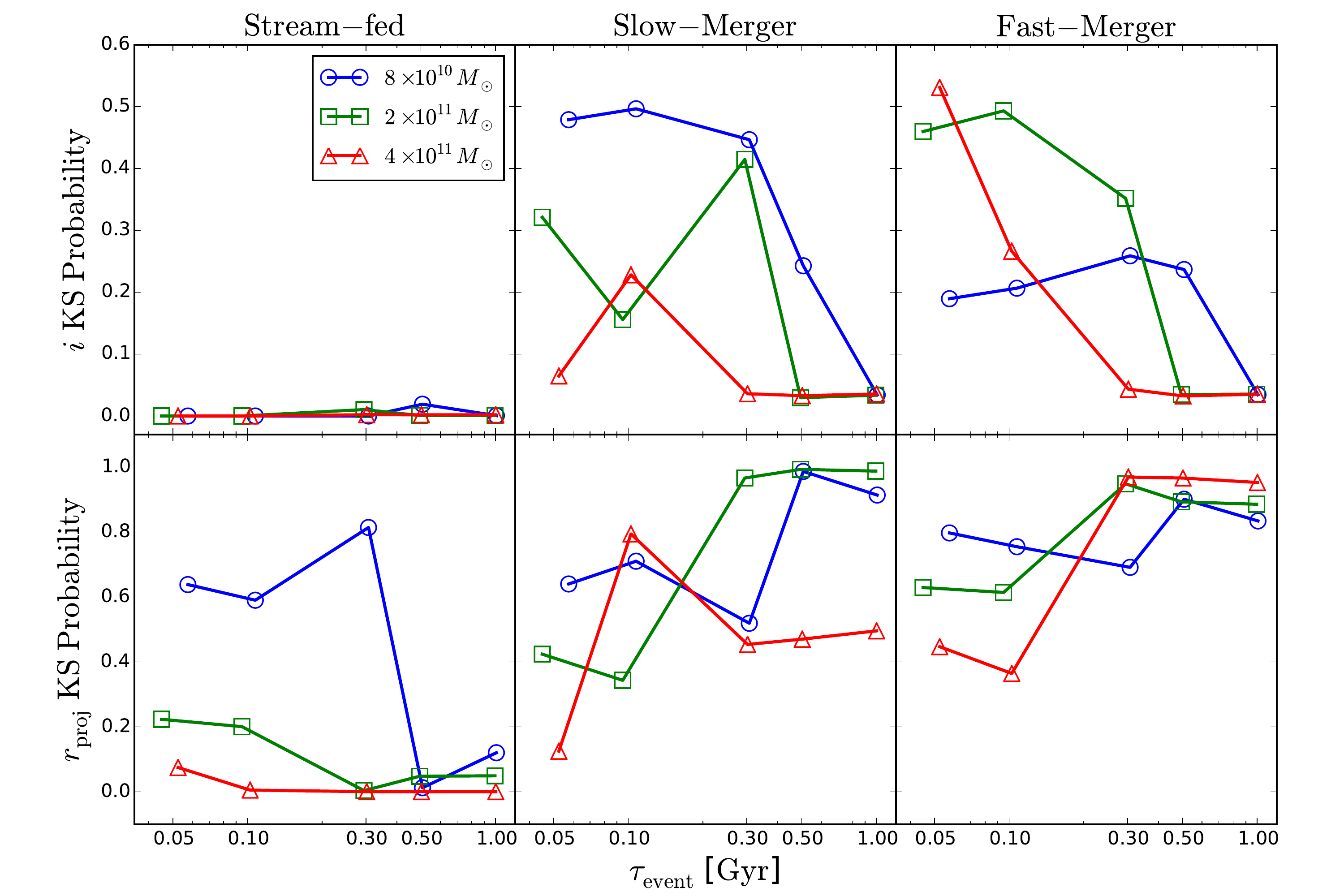}
 }
 \caption{KS model probabilities for the projected incidence angle $i$ (top row) and the projected radius $r_{\rm pro}$ (bottom row) for the sample of cluster members (green squares in Figs.~\ref{fig:incid_cum_dist} and \ref{fig:rad_cum_dist}), shown as a function of the duration of the RPS event, $\tau_\mathrm{event}$, and the mass of the respective galaxy (see legend). Infall along filaments (leftmost panels) is clearly disfavoured.\label{fig:ks_scores}}
\end{figure*}

\section{Results}

In order to reduce contamination by interlopers (fore- or background galaxies), we restrict our analysis to the subset of candidate RPS events with measured redshifts within $\pm$4000 km s$^{-1}$ of the redshift of the host cluster; the 53 objects (of 124 with measured redshifts) meeting this criterion are hereafter referred to as the ``spectroscopic sample". Of these, we select a subset of the 15 galaxies exhibiting the most compelling ``jellyfish" morphology comprised of the six systems presented by ESE and the nine shown in Fig.~\ref{fig:jellies} (``jellyfish sample"). We further restrict the comparison between data and model predictions to a projected radius of 415 kpc from the cluster core, which leaves 23 and 11 galaxies in the spectroscopic and jellyfish samples respectively. This radial cutoff minimises systematic incompleteness introduced at larger cluster-centric radii, which are covered only by images of the most distant clusters in our sample. 

Fig.~\ref{fig:incid_cum_dist} shows the cumulative distributions of the incidence angle for our two RPS subsamples plotted against predictions from our infall model, with (bottom row) and without (top row) correction for the bias discussed in Section~\ref{sec:bias} and \ref{sec:bias2}. The bottom The left, center, and right columns of Fig.~\ref{fig:incid_cum_dist} show predictions for the stream-fed, slow-merger, and fast-merger models, respectively (see Table~\ref{tab:models} for the parameters characterising these models).

Visual comparison suggests that the observations are best matched by the model predictions for the slow-merger scenario, provided that the duration of the stripping process is less than a Gyr\footnote{Note that, in the top panel of Fig.~\ref{fig:incid_cum_dist}, all of the solid lines, as well as the red dashed line, fall on top of each other and are thus indistinguishable by eye.}. Contrary to the traditional picture of RPS being driven purely by infall from the low-density field, preferably along filaments, we find poor agreement between the data and stream-fed models which over-predict events at extreme incidence angles (at $\lesssim$40$^\circ$ for almost all combinations of model parameters explored by us, and at $\gtrsim$140$^\circ$ for low-mass galaxies experiencing long RPS events).  In this scenario, the motion of galaxies is dominated by the cluster potential, which leads to a preferential alignment of trajectories toward the cluster core (at least in our projected view) and thus a highly anisotropic distribution of incidence angles. 
 
Fig.~\ref{fig:rad_cum_dist} shows the cumulative distributions of the number of RPS events within a given projected cluster-centric radius. To provide more natural, equal-area sampling, we bin the data in equal steps of $r_\mathrm{proj}^2$; a uniform areal distribution thus appears as a straight line from zero to one. We find that both of the cluster merger models predict a nearly uniform areal distribution of events in agreement with our observations. Stream-fed models with the most massive galaxies and/or the longest event timescales predict an excess at small projected radii which is not supported by our data. Note, that this comparison also effectively rules out the stream-fed model with a Milky-Way sized galaxy and 300 Myr timescale that at least marginally matched the observed distribution of incidence angles and is shown as the green dashed line in Fig.~\ref{fig:incid_cum_dist}.

A more quantitative assessment of the significance of the discrepancies between the observed and predicted distributions can be obtained with Kolmogorov-Smirnov (KS) tests. In Fig.~\ref{fig:ks_scores}, we show KS probabilities for the null hypothesis that the observed distributions are drawn from the same parent population as the predictions of a given model. Correcting all models for the aforementioned line-of-sight bias (Sections~\ref{sec:bias} and \ref{sec:bias2}) does not change our conclusions significantly. For simplicity, we therefore ignore the bias due to motion along the line of sight in the KS tests. To maximize the number of objects in the comparison, we show results for the spectroscopic sample only. However, considering the smaller jellyfish subsample does not significantly alter our conclusions.

Consistent with our qualitative assessments above,  we find no agreement with the observed distribution of incidence angles for any model assuming infall along filaments, although the distribution of projected radii does not rule out such models (at least not for low-mass galaxies, see bottom panel of Fig.~\ref{fig:ks_scores}). By contrast, practically all of the models for the two merger scenarios provide an acceptable (or good) description of the data, with the exception of those involving the most massive galaxies, for which models assuming long RPS durations of $\tau_\mathrm{event}\gtrsim$ 300 Myr are ruled out at more than $2\sigma$ confidence.

\section{Conclusions}

Since our models are intrinsically three dimensional, the comparisons presented above, although involving solely parameters measured in projection, allow us to distinguish between distinctly different three-dimensional scenarios. 

In the merger scenarios, RPS events are triggered in fast-moving galaxies near the outskirts of the cluster and, due to the relatively short duration of $\sim$500 Myr required by our incidence angle data (see top row of Fig.~\ref{fig:ks_scores}), remain confined to a shell well outside a (three-dimensional) cluster-centric radius of 400 kpc.  On the other hand, the projected radius data favour event durations longer than $\sim$100 Myr to explain the uniform areal distribution (Fig.~\ref{fig:rad_cum_dist}).  The RPS candidates detected by us are thus the projection of the essentially uniform distribution of much more distant RPS events in the fore- and background segments of this shell. In principle, galaxies of all masses may contribute to the observed RPS distribution; however, the majority are likely to be systems of low to intermediate mass, since models for extremely massive galaxies generally require finely tuned, short RPS lifetimes of about 100 Myr approximately to match the observations (red lines in Fig.~\ref{fig:ks_scores}).

By contrast, galaxies falling into the cluster along filaments do so at much lower peculiar velocities and thus require higher ICM densities for the GG criterion to be met; as a result,  RPS events are triggered only much closer to the cluster core.  To match the observed, broad distribution of incidence angles, these galaxies need time to enter our field of view from all sides, which mandates that the associated RPS events remain observable for 300 Myr or longer (Fig.~\ref{fig:incid_cum_dist}). Such long life-times, however, lead in turn to an excess in the number of events close to the cluster core that is not observed (Fig.~\ref{fig:rad_cum_dist}). 

We therefore tentatively conclude that extreme RPS events in massive clusters are generally short-lived ($\lesssim$500 Myr) and triggered far from the cluster core, likely driven by cluster mergers. Interestingly this preference of our analysis for RPS events being most readily observed in galaxies moving at high speed through an only modestly dense ICM suggests that textbook cases of ``jellyfish galaxies" might also be observed near the cores of less massive clusters (or even groups of galaxies, see also \citealt{poggianti_2015}) provided a cluster or group merger event ensures sufficiently high peculiar initial velocity. Note also that, while our data disfavour infall along filaments as the primary trigger, they do not  rule out a contribution from such a scenario. Wide-field imaging surveys that are able to detect RPS events out to the virial radius are needed to determine the relative contributions of stream-fed infall and cluster mergers.

\section{Summary}

We have conducted a systematic search for galaxies experiencing ram-pressure stripping (RPS) in 63 MACS clusters at $z{=}$0.3--0.7. Using quantitative morphological parameters for $\sim$16,000 galaxies detected in \emph{Hubble Space Telescope} images of these systems we identify 211 potential cases of RPS that complement a training set of 12 ``jellyfish" galaxies used to define our selection criteria.  Where possible, the direction of motion in the plane of the sky is estimated for these systems based on morphological indicators such as the curvature and orientation of the apparent galaxy-ICM interface region or a visible debris trail.  Several systematic biases are inherent to our approach: (a) the classification of galaxies according to their likelihood of undergoing RPS is partly based on visual inspection and thus to some extent subjective, (b) the small field of field of view our observations prevents us from sampling the galaxy population in the outer regions of our cluster targets (except in projection) where RPS events might be initially triggered, and (c) our selection process is fundamentally unable to robustly identify RPS events in galaxies moving along, or close to, our line of sight.

We attempt to address the first of these biases by obtaining spectroscopic redshifts of all our RPS candidates. While the resulting spectra do not immediately confirm or refute an RPS event, they allow us to establish whether or not a morphologically selected candidate is in fact a cluster member and whether its spectral characteristics are consistent with ongoing or recent star formation. So far, 53 of 124 systems targeted in spectroscopic follow-up observations were confirmed as cluster members. A detailed analysis of these galaxies' spectral properties will be presented in a forthcoming paper (Blumenthal et al., in preparation). 

The remaining two observational biases mentioned above can be accounted for by three-dimensional modelling of the trajectories and environment of galaxies falling into a massive cluster. Specifically, we
compare the distributions of the observed projected incidence angle and distance from the BCG with predictions from simple models of galaxy orbits in a MACS-like cluster.  We investigate two scenarios: accretion of galaxies from an attached filament, and a cluster merger event. 

We find significantly better agreement for the merger scenario, provided the duration of RPS events is $\lesssim$500 Myr.  We thus tentative conclude that extreme ram-pressure stripping events is primarily triggered in massive cluster mergers (rather than by infall alone) where relative velocities between galaxies and the ICM are large enough to initiate RPS far from the cluster core ($\gg$ 400 kpc). Although our study is, by design, limited to relatively massive clusters, we note that this result implies that extreme RPS events may also occur in mergers of poorer clusters and even groups of galaxies, where the required ingredients (high peculiar velocity and moderately high ICM density) are both met by galaxies close to core passage.
We also find that galaxies of mass similar to, or less than, our Milky Way are likely to dominate the set of observable RPS events in massive clusters, although more massive galaxies may contribute too at a lower level. Although models assuming infall along a filament were found to yield predictions that are largely in conflict with our data, both processes (accretion along filaments and via cluster mergers) can be expected to contribute.  The extent to which the two mechanisms are responsible for the observed population of RPS events in our sample is difficult to quantify but could be tested by imaging surveys that probe the distribution of RPS events to larger cluster-centric radii.


In-depth studies of the X-ray properties of RPS host clusters along with spectroscopic investigations of the star-formation rates and histories of the candidates identified in this study will be critical to test our conclusions and allow a quantitative comparison of observational diagnostics with predictions of numerical models of ram-pressure stripping.

\section*{Acknowledgements}
We thank the anonymous referee for their helpful comments, questions, and suggestions on revising the manuscript. CM thanks J. Lotz for providing the galaxy morphology source code which was adapted for this work. HE gratefully  acknowledges financial support from STScI grants GO-10495, -10875, -12166, and -12884.  This research made use of Astropy, a community-developed core Python package for Astronomy (Astropy Collaboration, 2013).

\appendix
\section{Skeletal Decomposition Parameters}
\label{sec:skel}

The morphological indicators discussed in Section~3 were generally defined to identify characteristic morphological traits of galaxy mergers \citep[e.g.][]{lotz_2011}.  We introduce a new metric based on the concept of the morphological skeleton \citep{maragos_1986} to both quantify the amount of substructure in a galaxy while concurrently identifying arm/tail-like structures.  Conceived in the context of mathematical morphology \citep[see ][]{serra1988image} and originally introduced as a means for binary image compression, the morphological skeleton (or medial axis transform) reduces a shape to a line that maintains the topological structure of the full image, thus allowing exact reconstruction.

We here generalise the definition of the morphological skeleton to images with non-binary, continuous greyscale pixel values. However, we must be cautious as noise in relatively short exposures used in this survey ($\sim$1200 sec.) can manifest as small scale substructure in the skeleton if applied naively. To reduce this erroneous signal from noise, we smooth the image using a Gaussian kernel before determining the skeleton and then prune the result to remove any disconnected segments. We define the result of this process as $Sk_i$. We perform skeletal decompositions under three smoothing scales corresponding to the Petrosian radius $r_p$, the half light radius $r_{50\%}$, and the 10\% light radius $r_{10\%}$ which define $Sk_0$, $Sk_1$, and $Sk_2$. Note that due to the cleaning process we apply here, exact reconstruction of the original image is not possible.

\begin{figure}
 \hspace*{-2mm}\includegraphics[width=0.5\textwidth]{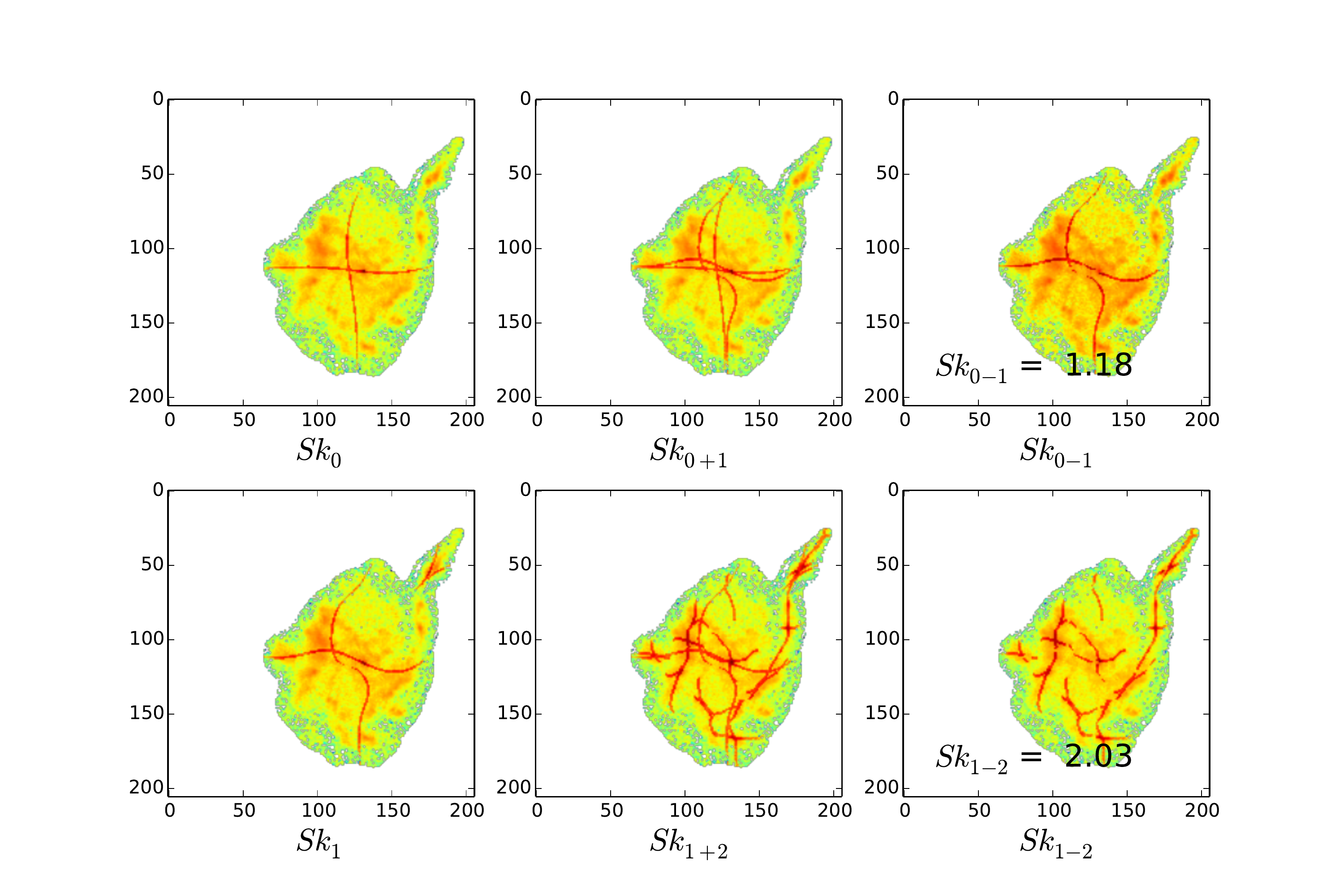}
 \caption{An example of the results of the greyscale skeletonization process which we use to define the skeletal decomposition parameters $Sk_{0-1}$ and $Sk_{1-2}$. \label{fig:Sk}}
 \end{figure}

To further reduce erroneous signal due to residual noise, we define $Sk_{x+y}$ (where $y=x+1$) as comprising all pixels in the higher-order skeleton (i.e. under a smaller smoothing kernel) connected to that of the lower-order skeleton (larger smoothing kernel). To generate a common reference point and to avoid bias due to image size and lower order structure, we then subtract the length of the lower-order skeleton from $Sk_{x+y}$ and normalise by the length of the lower-order skeleton (e.g. $[|Sk_{0+1}|-|Sk_0|]/|Sk_0|$) defining a final numerical measure $Sk_{x-y}$ which quantifies the excess in substructure under smoothing scale $y$ with respect to $x$ (see Fig.~\ref{fig:Sk}).

A simple way to understand this qualitatively is to consider a case where $Sk_{0-1}$ or $Sk_{0-1}$ is equal to zero. This would imply that image smoothed on a finer scale (smaller kernel) does not reveal any more substructure or that the galaxy's light profile is essentially smooth below the upper smoothing scale. However, as a full interpretation of the meaning and reliability of these indicators is beyond of the scope of this paper, we here characterise $Sk_{0-1}$ only to be a measure of bending in the galaxy or the deviation from a symmetric object (somewhat correlated with asymmetry), while $Sk_{1-2}$ quantifies the amount of clumpy substructure connected to the brighter regions of the galaxy.

\bibliography{jellies_paper}

\begin{table*}
  \caption{SNAPS Observations\label{tab:snaps_obs}}
  \begin{tabular}{lcccc}
    \hline 
    Name & $\alpha$ [J2000] & $\delta$ [J2000] & $t_{exp}$ [s] & GO Prop. ID \\ 
    \hline 
    \hline 
    EMACSJ1057.5+5759 & 10:57:31.680 & +57:59:33.72 & 1200 & 12884 \\ 
    MACSJ0032.1+1808 & 00:32:11.344 & +18:07:49.37 & 1200 & 12166 \\ 
    MACSJ0035.4-2015 & 00:35:26.957 & -20:15:50.66 & 1200 & 10491 \\ 
    MACSJ0140.0-0555 & 01:40:01.626 & -05:55:06.71 & 1200 & 10491 \\ 
    MACSJ0152.5-2852 & 01:52:35.361 & -28:53:39.88 & 1200 & 10491 \\ 
    MACSJ0257.6-2209 & 02:57:40.596 & -22:09:27.80 & 1200 & 10875 \\ 
    MACSJ0308.9+2645 & 03:08:56.839 & +26:45:43.91 & 1200 & 12166 \\ 
    MACSJ0451.9+0006 & 04:51:55.443 & +00:06:11.66 & 1200 & 10491 \\ 
    MACSJ0521.4-2754 & 05:21:25.808 & -27:55:06.91 & 1200 & 10491 \\ 
    MACSJ0547.0-3904 & 05:47:01.796 & -39:04:13.24 & 1200 & 12166 \\ 
    MACSJ0553.4-3342 & 05:53:23.850 & -33:42:42.21 & 2092 & 12362 \\ 
    MACSJ0712.3+5931 & 07:12:21.985 & +59:32:24.82 & 1200 & 10491 \\ 
    MACSJ0845.4+0327 & 08:45:28.224 & +03:27:28.46 & 1200 & 10491 \\ 
    MACSJ0916.1-0023 & 09:16:12.344 & -00:23:47.00 & 1200 & 10491 \\ 
    MACSJ0947.2+7623 & 09:47:10.744 & +76:23:21.62 & 1200 & 10491 \\ 
    MACSJ0949.8+1708 & 09:49:52.655 & +17:07:06.38 & 1200 & 10491 \\ 
    MACSJ1006.9+3200 & 10:06:55.632 & +32:01:33.91 & 1200 & 10491 \\ 
    MACSJ1115.2+5320 & 11:15:15.968 & +53:19:47.47 & 1200 & 10491 \\ 
    MACSJ1124.5+4351 & 11:24:29.365 & +43:51:32.97 & 1200 & 12166 \\ 
    MACSJ1133.2+5008 & 11:33:14.109 & +50:08:29.50 & 1200 & 10491 \\ 
    MACSJ1142.4+5831 & 11:42:26.434 & +58:32:01.30 & 1200 & 12166 \\ 
    MACSJ1226.8+2153C & 12:26:41.421 & +21:53:07.58 & 1200 & 12166 \\ 
    MACSJ1236.9+6311 & 12:36:59.868 & +63:11:02.26 & 1200 & 10491 \\ 
    MACSJ1258.0+4702 & 12:58:02.708 & +47:02:42.87 & 1200 & 10491 \\ 
    MACSJ1319.9+7003 & 13:20:09.685 & +70:04:28.16 & 1200 & 10491 \\ 
    MACSJ1354.6+7715 & 13:54:31.253 & +77:15:08.71 & 1200 & 10491 \\ 
    MACSJ1447.4+0827 & 14:47:26.289 & +08:28:37.08 & 1200 & 12166 \\ 
    MACSJ1452.9+5802 & 14:52:57.957 & +58:02:43.28 & 1200 & 12166 \\ 
    MACSJ1526.7+1647 & 15:26:42.342 & +16:47:48.83 & 1200 & 12166 \\ 
    MACSJ1621.3+3810 & 16:21:23.928 & +38:10:16.28 & 1200 & 12166 \\ 
    MACSJ1644.9+0139 & 16:45:01.729 & +01:40:09.83 & 1200 & 12166 \\ 
    MACSJ1652.3+5534 & 16:52:19.726 & +55:34:46.63 & 1200 & 10491 \\ 
    MACSJ1731.6+2252 & 17:31:39.268 & +22:52:05.09 & 1200 & 12166 \\ 
    MACSJ1738.1+6006 & 17:38:05.383 & +60:06:14.92 & 1200 & 12166 \\ 
    MACSJ1752.0+4440 & 17:51:57.961 & +44:39:45.45 & 1200 & 12166 \\ 
    MACSJ1806.8+2931 & 18:06:51.898 & +29:30:23.03 & 1200 & 12166 \\ 
    MACSJ2050.7+0123 & 20:50:42.381 & +01:23:24.69 & 1200 & 12166 \\ 
    MACSJ2051.1+0215 & 20:51:10.058 & +02:16:00.72 & 1200 & 12166 \\ 
    MACSJ2135.2-0102 & 21:35:12.822 & -01:02:51.52 & 1200 & 10491 \\ 
    MACSJ2241.8+1732 & 22:41:56.386 & +17:32:47.33 & 1200 & 12166 \\ 
    SMACSJ0234.7-5831 & 02:34:43.512 & -58:31:16.51 & 1200 & 12166 \\ 
    SMACSJ0549.3-6205 & 05:49:18.358 & -62:05:07.88 & 1200 & 12166 \\ 
    SMACSJ0600.2-4353 & 06:00:12.915 & -43:53:19.33 & 1200 & 12166 \\ 
    SMACSJ0723.3-7327 & 07:23:18.709 & -73:27:06.01 & 1200 & 12166 \\ 
    SMACSJ2031.8-4036 & 20:31:46.993 & -40:37:03.68 & 1200 & 12166 \\ 
    SMACSJ2131.1-4019 & 21:31:05.693 & -40:19:12.22 & 1200 & 12166 \\ 
    \hline 
  \end{tabular}
\end{table*}

\begin{table*}
  \caption{CLASH Observations\label{tab:clash_obs}}
  \begin{tabular}{lcccc}
    \hline 
    Name & $\alpha$ [J2000] & $\delta$ [J2000] & $t_{exp}$ [s] & GO Prop. ID \\ 
    \hline 
    \hline 
    MACSJ0329-0211 & 03:29:41.560 & -02:11:46.10 & 4104 & 12452 \\ 
    MACSJ0416-2403 & 04:16:08.380 & -24:04:20.79 & 4036 & 12459 \\ 
    MACSJ0429-0253 & 04:29:36.049 & -02:53:06.10 & 3938 & 12788 \\ 
    MACSJ0647+7015 & 06:47:50.269 & +70:14:54.99 & 4128 & 12101 \\ 
    MACSJ0717.5+3745-POS5 & 07:17:32.629 & +37:44:59.70 & 7920 & 10420 \\ 
    MACSJ0744+3927 & 07:44:52.819 & +39:27:26.89 & 4128 & 12067 \\ 
    MACSJ1115+0129 & 11:15:51.900 & +01:29:55.10 & 3870 & 12453 \\ 
    MACSJ1149+2223 & 11:49:34.704 & +22:24:04.75 & 4128 & 12068 \\ 
    MACSJJ1206.2-0847 & 12:06:12.055 & -08:47:59.44 & 6608 & 10491 \\ 
    MACSJ1311-0310 & 13:11:01.800 & -03:10:39.79 & 4158 & 12789 \\ 
    RXJ1347-1145 & 13:47:32.110 & -11:45:11.36 & 3878 & 12104 \\ 
    MACS1423+2404 & 14:23:47.88 & +24:04:42.49 & 4240 & 12790 \\ 
    RXJ1532+3021 & 15:32:53.779 & +30:20:59.39 & 4060 & 12454 \\ 
    MACSJ1720+3536 & 17:20:16.780 & +35:36:26.49 & 4040 & 12455 \\ 
    MACSJ1931-2635 & 19:31:49.62 & -26:34:32.90 & 3850 & 12456 \\ 
    MACSJ2129-0741 & 21:29:26.059 & -07:41:28.79 & 3728 & 12100 \\ 
    RXJ2248-4431 & 22:48:43.960 & -44:31:51.30 & 3976 & 12458 \\ 
    \hline 
  \end{tabular}
\end{table*}

\end{document}